\documentclass[aps,prd,twocolumn,groupedaddress,preprintnumbers,showpacs,showkeys,tightenlines,superscriptaddress,nofootinbib,nobibnotes]{revtex4-2}
\usepackage{lipsum}
\usepackage{hyperref}
\usepackage{epsfig}
\usepackage{latexsym}
\usepackage{amssymb}
\usepackage{booktabs}
\usepackage{graphicx}\usepackage{physics}
\usepackage{simpler-wick}

\usepackage{xcolor}
\usepackage{amsmath}
\usepackage{bm}

\let\oldeqref\eqref
\renewcommand{\eqref}[1]{Eq.~\oldeqref{#1}}

\newcommand{\figref}[1]{Fig.~\ref{#1}}
\newcommand{\tabref}[1]{Tab.~\ref{#1}}

\newcommand{\secref}[1]{Sec.~\ref{#1}}

\begin{document}
\preprint{MITP-26-022}
\title{The strange and flavor-singlet axial form factors of the nucleon from lattice QCD}

\author{Alessandro Barone}
\email{abarone@uni-mainz.de}
\affiliation{PRISMA$^++$ Cluster of Excellence \& Institut f\"ur Kernphysik,
 Johannes Gutenberg-Universit\"at  Mainz,  D-55099 Mainz, Germany}
 
\author{Dalibor~Djukanovic} 
\affiliation{Helmholtz-Institut Mainz, Johannes Gutenberg-Universit\"at Mainz,
D-55099 Mainz, Germany}
 \affiliation{GSI Helmholtzzentrum für Schwerionenforschung, 64291 Darmstadt, Germany}

\author{Georg~von~Hippel}
\affiliation{PRISMA$^++$ Cluster of Excellence \& Institut f\"ur Kernphysik,
 Johannes Gutenberg-Universit\"at  Mainz,  D-55099 Mainz, Germany}

\author{Harvey~B.~Meyer} 
\affiliation{PRISMA$^++$ Cluster of Excellence  \& Institut f\"ur Kernphysik,
Johannes Gutenberg-Universit\"at Mainz, D-55099 Mainz, Germany}
\affiliation{Helmholtz-Institut Mainz, Johannes Gutenberg-Universit\"at Mainz,
D-55099 Mainz, Germany}
\affiliation{Theoretical Physics Department, CERN, 1211 Geneva 23, Switzerland}

\author{Konstantin~Ottnad}
\affiliation{PRISMA$^++$ Cluster of Excellence \& Institut f\"ur Kernphysik,
 Johannes Gutenberg-Universit\"at  Mainz,  D-55099 Mainz, Germany}
\affiliation{Helmholtz-Institut f\"ur Strahlen- und Kernphysik, Universit\"at Bonn, D-53115 Bonn, Germany}

\author{Hartmut~Wittig} 
 \affiliation{PRISMA$^++$ Cluster of Excellence  \& Institut f\"ur Kernphysik,
Johannes Gutenberg-Universit\"at Mainz, D-55099 Mainz, Germany}
\affiliation{Helmholtz-Institut Mainz, Johannes Gutenberg-Universit\"at Mainz,
D-55099 Mainz, Germany}

\begin{abstract}

The singlet axial form factor of the nucleon provides essential input for
a complete understanding of the nucleon axial structure.
Together with the isovector and isoscalar octet channels, in the forward limit it forms the basis
for a full flavor decomposition of the proton spin.
In this work we present a lattice QCD determination of the singlet axial form factor $G^{u+d+s}_A(Q^2)$
and related strange contribution $G^{s}_A(Q^2)$
using a set of $N_f = 2 + 1$ CLS gauge ensembles with $O(a)$-improved Wilson fermions,
with a full error budget for the extrapolation to the chiral, continuum and infinite-volume limits.
Particular focus is placed on the treatment of the disconnected contributions,
which constitute the crucial element for the extraction of the strange component.
Together with determinations of the isovector and isoscalar octet
axial form factors, this work
provides a comprehensive lattice QCD determination of the nucleon
axial structure across different flavor channels.

\end{abstract}

\maketitle

\section{Introduction\label{sec:intro}}

The axial form factor of the nucleon plays a central role in our understanding of
the electroweak interactions of nucleons as well as the proton's intrinsic structure. 
The axial form factor enters a wide range of processes such as neutron $\beta$ decay and neutrino-nucleon scattering,
contributing to our theoretical understanding of neutrino cross sections
relevant for current and future long-baseline neutrino-oscillation experiments.
In addition, charges of the axial current encode information on the fraction of the proton spin carried
by quarks~\cite{Ji:1996ek,COMPASS:2006mhr,Aidala:2012mv}.

In the SU(3) flavor basis, the nucleon axial current can be decomposed into isovector $(u-d)$,
isoscalar octet $(u+d-2s)$, and singlet $(u+d+s)$ components.
These channels probe distinct aspects of QCD:
the isovector channel is sensitive to $W$ boson exchange, whereas
the isoscalar currents are sensitive to $Z$ boson scattering, which also receives contributions from
the strange quark.
In particular, the isovector channel is relevant
in the analysis of upcoming neutrino experiments~\cite{DUNE:2015lol,Hyper-Kamiokande:2018ofw}, 
whereas channels involving $Z$ boson exchange and sea-quark contributions are needed
as input for the precise extraction of the weak charge of the proton in the upcoming P2 experiment
\cite{Becker:2018ggl}. Furthermore, the strange component $G_A^{s}(Q^2)$ is
crucial for calculating the cross-section within various dark matter models~\cite{Papavassiliou:2009zz}:
in this context, it is particularly relevant for the MicroBooNE neutrino detector at Fermilab, which aims
to extract $G_A^{s}(Q^2)$ in the range from $Q^2 = 0.08 \,\text{GeV}^2$ to $Q^2 = 1 \,\text{GeV}^2$~\cite{Miceli:2014hva,Kim:2019mdm}.

Lattice QCD provides a non-perturbative framework for computing nucleon axial form factors from first principles.
While the isovector combination has seen substantial progress~\cite{RQCD:2019jai,Hasan:2019noy,Alexandrou:2020okk,Gupta:2017dwj,
Djukanovic:2022wru,Jang:2019vkm,Tomalak:2023pdi,Jang:2023zts,Meyer:2026kdl},
the remaining flavor channels have so far been explored less extensively at the
level of full form-factor calculations. Existing studies of the singlet and
strange channels~\cite{Green:2017keo,Alexandrou:2021wzv} do not include a
controlled continuum extrapolation, while several lattice determinations of the
corresponding axial charges are available~\cite{Liang:2018pis,Lin:2018obj,Park:2020axe,Alexandrou:2024ozj}.
The first complete lattice determination of the isoscalar octet channel has
been reported only recently~\cite{Barone:2025hhf,Barone:2025rye}.
Indeed, while the $(u-d)$ combination is free of quark-disconnected contributions, these
appear in the isoscalar channels, requiring extra computational cost and a more careful analysis.
Moreover, in the singlet case, the axial current is further distinguished by the presence of the axial anomaly,
which introduces gluonic contributions leading to additional complications in both the
lattice calculation and the renormalization of the corresponding matrix elements.

In a series of previous works~\cite{Djukanovic:2022wru,Barone:2025rye}, we have investigated the nucleon axial form factor
using a set of $N_f=2+1$ CLS gauge ensembles and a common analysis strategy.
Our earlier publications presented results for the isovector axial form factor~\cite{Djukanovic:2022wru}
and for the isoscalar octet combination~\cite{Barone:2025rye}.
In the present work, we extend this program to the SU(3)$_f$ singlet axial form
factor and, simultaneously, to the strange axial form factor.

Our goal in this work is to determine the momentum-transfer
dependence, highlighting the impact of disconnected contributions.
In particular, we focus on the treatment of the strange-quark component,
which is combined with the previously determined octet contribution to construct the flavor-singlet channel.
By completing the analysis of the isovector, octet, and singlet axial form factors within a
single computational framework, this work provides a comprehensive lattice QCD determination
of the channels relevant for the quark-flavor decomposition of the nucleon axial structure in the SU(3)
flavor basis.

This paper is organized as follows. In Sec.~\ref{sec:setup} we briefly summarize the lattice setup
and correlation-function construction common to the previous works, and we emphasize 
aspects specific to the singlet axial current.
Section~\ref{sec:analysis} describes the analysis strategy and treatment of disconnected contributions specific to this work, 
and the extraction of the strange and singlet form factors.
We present our results for the form factors in Sec.~\ref{sec:results},
and conclude in Sec.~\ref{sec:conclusions}.

\section{Lattice setup\label{sec:setup}}

We consider the axial-vector currents
\begin{align}
 A_\mu^a = \bar \psi (x) \gamma_\mu\gamma_5 \lambda^a \psi(x)\,, \quad \psi=(u,d,s)^T \,,
\end{align}
where $\lambda^a$ are the Gell-Mann matrices, $\lambda^0 = \sqrt{2/3}\, \mathbb{I} $, and explicitly
\begin{align}
 A_\mu^0 &\equiv \sqrt{\frac{2}{3}} A_\mu^{u+d+s} = \sqrt{\frac{2}{3}}\left(\bar{u}\gamma_\mu\gamma_5 u + \bar{d}\gamma_\mu\gamma_5 d + \bar{s}\gamma_\mu\gamma_5 s \right) \,, \\  \notag
 A_\mu^8 &\equiv \frac{1}{\sqrt{3}} A_\mu^{u+d-2s} = 
 \frac{1}{\sqrt{3}} \left( \bar{u}\gamma_\mu\gamma_5 u + \bar{d}\gamma_\mu\gamma_5 d -2 \bar{s}\gamma_\mu\gamma_5 s \right) \, .
\end{align}
We use the following parameterization of their matrix elements
\begin{align}
\label{eq:matrix_el}
 \langle N_{\rm p^+} | A_\mu^a |N_{\rm p^+}\rangle& = \bar u (p')\Bigl [ G_A^a(Q^2) \gamma_\mu + \frac{q_\mu}{2m} G_P^a(Q^2)\Bigr] \gamma_5\,u(p)
\end{align} 
where $ N_{\rm p^+}$ indicates the proton state,
$p$ $(p')$ is the momentum of the incoming (outgoing) proton, $q=p'-p$
the momentum transfer, $p^2=p'{}^2=m^2$ and $q^2= -Q^2$.

The lattice calculation relies on the same data used in our previous works~\cite{Djukanovic:2022wru,Barone:2025rye}, i.e.
fourteen CLS~\cite{Bruno:2014jqa} $N_f=2+1$ gauge ensembles
with $\mathcal{O}(a)$ improved Wilson fermions \cite{Sheikholeslami:1985ij}
and tree-level improved L{\"u}scher-Weisz
gauge-action \cite{Luscher:1984xn}. They span a range of lattice 
spacings from $0.050\,\text{fm}$ to $0.086\,\text{fm}$ and pion masses
from $130\,\text{MeV}$ to $350\,\text{MeV}$. 
The necessary reweighting factors to correct for the treatment of the strange-quark
determinant have been computed according to~\cite{Mohler:2020txx,Kuberski:2023zky}.
A brief summary of the ensembles employed in this study is provided in Table~\ref{tab:ensembles}, and we refer to~\cite{Djukanovic:2022wru}
for further details. 

\begin{table}[t]
	\begin{tabular}{cccccc}
\toprule
		ID & $\beta$ & $T/a$ & $L/a$ & $M_\pi$ [MeV] & $M_N$ [MeV]
		\\\hline
		H102 & 3.40 & 96 & 32 & 354 & 1103 \\
		H105 & 3.40 & 96 & 32 & 280 & 1045 \\
		C101 & 3.40 & 96 & 48 & 225 & 980 \\
		N101 & 3.40 & 128 & 48 & 281 & 1030\\\hline
		S400 & 3.46 & 128 & 32 & 350 & 1130 \\
		N451 & 3.46 & 128 & 48 & 286 & 1011\\
		D450 & 3.46 & 128 & 64 & 216 & 978\\ \hline
		N203 & 3.55 & 128 & 48 & 346 & 1112\\
		N200 & 3.55 & 128 & 48 & 281 & 1063\\
		D200 & 3.55 & 128 & 64 & 203 & 966 \\
		E250 & 3.55 & 192 & 96 & 129 & 928 \\\hline
		N302 & 3.70 & 128 & 48 & 348 & 1146 \\
		J303 & 3.70 & 192 & 64 & 260 & 1048 \\
		E300 & 3.70 & 192 & 96 & 174 & 962 \\
		\bottomrule
	\end{tabular}
	\caption{Summary of ensembles used, where the values for
	$\beta=3.40,3.46,3.55,3.70$ correspond to a lattice spacing of roughly
	$a\sim 0.086,0.076,0.064,0.050$ fm, respectively~\cite{Strassberger:2021tsu,RQCD:2022xux}. }
	\label{tab:ensembles}
\end{table}

\subsection{Correlation Functions}

The setup, conventions, and choice of spin projectors follow those adopted in our previous analyses,
which we briefly recall here.
For the computation of the relevant correlation functions we employ the nucleon interpolating field
\begin{align}
\label{eq:interp}
  \Psi_\alpha (x) = \varepsilon_{abc} \Bigl( \tilde{u} _a^T (x) C \gamma_5
  \tilde{d}_b(x) \Bigr) \tilde{u}_{c,\alpha} (x),
\end{align}
using Gaussian smeared quark fields $\tilde{q}(x)$~\cite{Gusken:1989ad}
and spatially APE-smeared gauge links in the covariant Laplacian~\cite{APE:1987ehd}.

We build the two-point correlation functions as
\begin{align}
	C_2^\Gamma(t,\bf{p})&=\Gamma_{\alpha \beta} \sum\limits_{\bf{x}} e^{-i {\bf p \cdot x }}
	\langle \Psi_\beta(\mathbf{x},t) \overline{\Psi}_\alpha (0)\rangle
\end{align}
as well as the connected three-point functions
\begin{align}
\label{eq:C3pt_conn}
 C^{\mathrm{conn},\Gamma'}_3(t,t_{s},{\bf q}) \hspace{-0.2em} =&  \Gamma'_{\alpha\beta} 
 \sum\limits_{\bf x, y} e^{i {\bf q \cdot y}} 
\frac{\mathbf{q}\cross \mathbf{s}}{|\mathbf{\bf q}\cross \mathbf{s}|^2} \,\cdot \\ \notag
	&\langle \Psi_\beta(\mathbf{x},t_s) \,\mathbf{q}\cross \mathbf{A} \, (\mathbf{y},t)\,
	\overline{\Psi}_\alpha (0)\rangle,
\end{align}
where the axial current is projected onto the orthogonal component to isolate the corresponding axial form factor,
cf.~\eqref{eq:matrix_el}.
In practice, we align the nucleon spin along the $z$-axis $\mathbf{s}=\mathbf{e}_3$,
and choose  the polarization matrix  $\Gamma' =\frac{1}{2} (1+\gamma_0) (1+i \gamma_5 \gamma_3)$.
Momentum is injected at the operator insertion,
and the sink is projected to zero momentum.

The disconnected contributions to the three-point correlator read
\begin{align}
\label{eq:C3pt_disc}
 C_{3,\mu}^{\mathrm{disc},\Gamma'} (t,t_s,{\bf q}) &=
\langle \mathcal{L}_\mu (t,{\bf q })\ \mathcal{C}^{\Gamma'}_2(t_{s},{\bf
	0})\rangle,
\end{align}
where $ \mathcal{L}_\mu$ indicates the one-point function
\begin{align}
	\mathcal L_\mu (z_0,{\bf q})&= -\sum\limits_{\bf z} e^{i {\bf q\cdot z }
	} \ \mathrm{Tr} \Bigl[ S_q^{-1}(z,z) \gamma_\mu \gamma_5\Bigr],
\end{align}
$S_q^{-1}$ being the propagator for the quark $q$,
and $\mathcal{C}_2$ denotes the Wick contraction of the interpolating operators
of Eq.~(\ref{eq:interp}),
\begin{align}
	\mathcal{C}^{\Gamma'}_2 (t_s,{\bf p'}) &= \sum_{\bf x} e^{-i {\bf p'\cdot  x}} \, 
	\text{Tr} \Bigl[ \Gamma' \cdot \wick{\c1 \Psi(\mathbf{x},t_s) \, {\overline{\c1
	\Psi}(0)} }
	\Bigr],
\end{align}
where the traces are in Dirac space.
In this case
we average over all three different polarizations, i.e. 
\begin{align}
	\Gamma'_i= \frac{1}{2}(1+\gamma_0) (1+ i \gamma_5 \gamma_i),\qquad
	i=1,2,3.
\end{align}
For the computation of the disconnected contributions
we use the same techniques applied in~\cite{Agadjanov:2023efe} (see also Appendix C of~\cite{Ce:2022eix} for full details),
namely stochastic estimates of the quark loops with variance reduction.

\subsection{Ratios and matrix elements}

Matrix elements are extracted from ratios of three- and two-point correlation functions designed to cancel
overlap factors and the leading Euclidean time dependence in the large-time limit.
The ratio is given by
\begin{align}
	R(t,t_s,{\bf q} )& =\frac{C^{\Gamma'}_3(t,t_s,{\bf
	q})}{C^{\Gamma}_2(t_s,{\bf 0})} \nonumber \\
	&\times \sqrt{
		\frac{C^{\Gamma}_2(t_s -t ,{\bf q}) C^{\Gamma}_2 (t,{\bf{0}})
		C^{\Gamma}_2(t_s,{\bf 0})}{C^{\Gamma}_2(t_{s}-t,{\bf 0})
		C^{\Gamma}_2(t,{\bf q}) C^{\Gamma}_2(t_s,{\bf q})} 
	} \, .
\label{eq:ratio}
\end{align}
In the connected case, the three-point function in~\eqref{eq:C3pt_conn} automatically projects the ratio onto the relevant
effective form factor, i.e.
\begin{align}
G_A^{\mathrm{conn,eff}}(t, t_s, {\bf q}) = \sqrt{\frac{2E_{\mathbf{q}}}{m+E_{\mathbf{q}}}}   R^{\rm conn}(t,t_s,{\bf q} )\, .
\end{align}

For the disconnected contribution, by contrast,
we follow the strategy described in~\cite{Barone:2025rye} to isolate the signal for $G_A$,
that we extract by solving the corresponding system of equations.
In this case, we use two-point
functions with higher statistics than in the connected ratio $R^{\rm conn}$,
which relies on the correlations between two-point and connected three-point
functions.
The final form factor is then given by $G_A^{\mathrm{eff}}=G_A^{\mathrm{conn,eff}}+G_A^{\mathrm{disc,eff}}$,
which in the large source-sink separation limit gives
\begin{align}
  G_A^{\mathrm{eff}}(t, t_s, {\bf q}) \xrightarrow{t_s - t,\, t\, \gg \Delta E^{-1}}  G_A(Q^2) \, ,
\end{align}
where $\Delta E$ stands for the energy gap in the one-nucleon sector.

\subsection{Renormalization}

The octet current used in~\cite{Barone:2025rye} makes use of the
$O(a)$-improved and mass-dependent renormalization factor
\begin{align}
\label{eq:renorm_8}
	A^{8,I}_{\mu,R} &= Z_A\Bigl[\bigl(1+ 3\overline{b}_A \, a m_q^{\rm av}   +\frac{b_A}{3} \,
	a (m_{l} + 2 m_{s})\bigr) A^{8,I}_\mu \nonumber\\
	&+ \left(\frac{1}{3} b_A+ f_A\right) \sqrt{2}
	 a (m_{l} -m_{s}) A^{0,I}_\mu\Bigr],
\end{align}
with $m_q^{\rm av} = (2 m_{l}+m_{s})/3$,
where we use the improvement coefficients for $c_A$, $b_A$ from
\cite{Bulava:2015bxa,Bali:2021qem},  and the  renormalization constant of the axial-vector current  $Z_A$ from
\cite{DallaBrida:2018tpn}.
The coefficient $\bar b_A$ corresponds to a genuine sea-quark effect and has been found to be small~\cite{Bali:2023sdi}; therefore, we neglect it.
We treat $f_A$ analogously,
as its contribution is expected to be numerically very small~\cite{Bhattacharya:2005rb}.
For the singlet case, on the other hand, we determine 
the renormalization factors as part of this work. We compute the
mass-independent renormalization factor on a set of $N_f=3$
ensembles, and report the details in the Appendix. We note that this is sufficient
for the current level of precision, due to the
statistical noise carried by the disconnected contributions.

The renormalized currents read
\begin{align}
\label{eq:renorm_matrix}
\begin{pmatrix}
A^3_{\mu,R} \\
A^8_{\mu,R} \\
A^0_{\mu,R}
\end{pmatrix}
=
\begin{pmatrix}
Z^{33}_A & 0 & 0 \\
0 & Z^{88}_A & Z^{80}_A  \\
0 & Z^{08}_A & Z^{00}_A
\end{pmatrix}
\begin{pmatrix}
A^3_{\mu} \\
A^8_{\mu} \\
A^0_{\mu}
\end{pmatrix} \,,
\end{align}
where we defined (cf.~\eqref{eq:renorm_8})
\begin{align}
 Z^{88}_A &= Z_A\left[ 1+ 3\overline{b}_A \, a m_q^{\rm av}   +\frac{b_A}{3} \, a (m_{l} + 2 m_{s}) \right] \,, \\ 
 Z^{80}_A &= Z_A\left[ \sqrt{2} \left(\frac{1}{3} b_A+ f_A\right)  a (m_{l} -m_{s})   \right] \,.
\end{align}
Since we determine only the mass-independent singlet renormalization factor,
we neglect the off-diagonal coefficient $Z_A^{08}$. This coefficient is induced
only by SU(3)-breaking $O(a)$ effects and hence scales as
$Z_A^{08}=O\!\left(a(m_l-m_s)\right)$, vanishing in both the continuum
and SU(3)-symmetric limits. At the present level of precision this contribution
is neglected. The calculation of $Z_A^{00}$ is described in Appendix~A, and the
corresponding values are listed in~\tabref{tab_ren_consts}.
The final results are reported in the $\overline{\text{MS}}$ scheme at 2 GeV.

It is then clear from~\eqref{eq:renorm_matrix} that the flavor-decomposed currents mix under renormalization.
In particular, we rewrite for convenience the renormalized isoscalar currents in terms of the bare light and strange components as
\begin{align}
 A^{u+d-2s}_{\mu,R} &\equiv Z_A^{u+d-2s, u+d}A^{u+d}_\mu -2 Z_A^{u+d-2s, s}A^{s}_\mu  \,, \\
 A^{u+d+s}_{\mu,R} &\equiv Z_A^{u+d+s, u+d}A^{u+d}_\mu + Z_A^{u+d+s, s}A^{s}_\mu  \,, 
\end{align}
where we defined
\begin{align}
 Z_A^{u+d-2s, u+d} &= Z_A\left( 1+b_A a m_{l}\right) \,, \\
 Z_A^{u+d-2s, s} &= Z_A\left( 1+b_A a m_{s}\right) \,, \\
 Z_A^{u+d+s,u+d} &= Z_A^{u+d+s,s} = Z_A^{00}.
\end{align}
For the renormalized strange-quark current we then obtain
\begin{align}
\label{eq:AsR}
 A^{s}_{\mu,R}
 & = \frac{1}{3}\left[ A^{u+d+s}_{\mu,R} - A^{u+d-2s}_{\mu,R} \right]  \\ \notag
 & \equiv Z_A^{s,\,u+d}\, A^{u+d}_{\mu}
   + Z_A^{s,\,s}\, A^{s}_{\mu} \, ,
\end{align}
where we have defined 
\begin{align}
\label{eq:ZAs_conn}
 Z_A^{s,\,u+d} &= \frac{1}{3}\left[ Z_A^{u+d+s,u+d} - Z_A^{u+d-2s,u+d} \right] \, , \\
 Z_A^{s,\,s}   &= \frac{1}{3}\left[ Z_A^{u+d+s,s} + 2 Z_A^{u+d-2s,s}\right] \, .
\end{align}
As a result, a mixing with the $u+d$ channel is present. Note, however, that this contribution
is numerically small in our calculation,
because $Z_A^{u+d+s,u+d} \simeq Z_A^{u+d-2s,u+d}$.

\section{Analysis method\label{sec:analysis}}

\begin{figure}[t]
\centering
\includegraphics[scale=0.3]{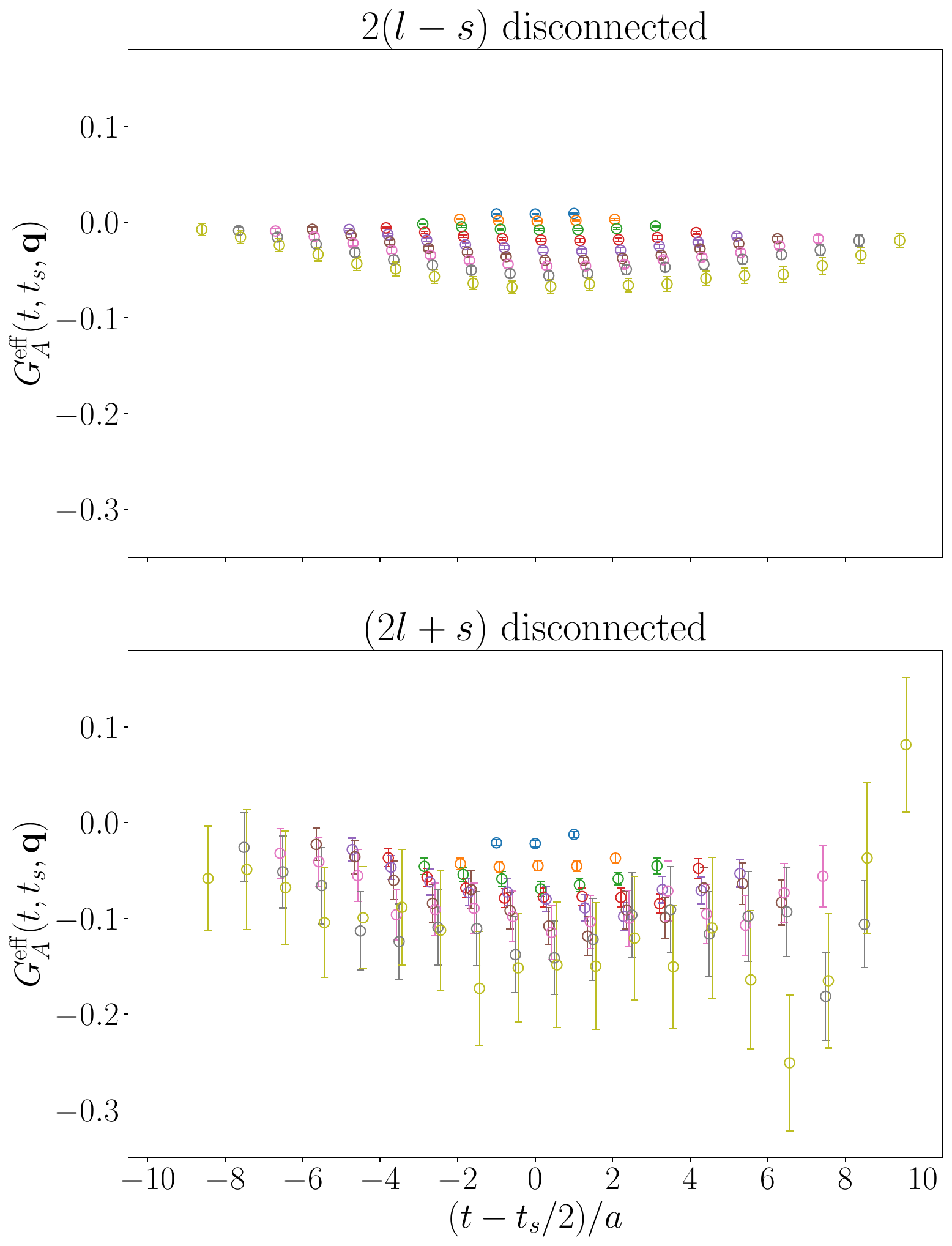}
\caption{Comparison between the disconnected contribution
to the isoscalar octet channel (top) and singlet (bottom) 
on D200 at first non-zero momentum $Q^2\simeq 0.089\, \text{GeV}^2$. The color indicates the different source-sink separations $t_s$.}
\label{fig:disc_octet_VS_singlet}
\end{figure}

In this section we describe the analysis strategy used to extract the strange and singlet axial form factor
from the lattice correlation functions. The treatment follows the methodology employed
in our previous work on the isoscalar octet, with modifications dictated by the presence of disconnected contributions 
that exhibit a substantial signal-to-noise problem,
since they do not benefit from a cancellation of ultraviolet noise as in the octet case $2(l-s)$. We show an
example of this for the ensemble D200 in~\figref{fig:disc_octet_VS_singlet}.
For this reason, connected and disconnected contributions are analyzed
separately using different strategies, and are subsequently combined in the determination of the physical form factor.

\subsection{Connected Contribution}
\label{subsec:connected}

To determine the connected contribution to the form factor
we make use of the summation method~\cite{Maiani:1987by,Capitani:2012gj}
to address the excited-state contamination and perform a linear fit to the summed expression
\begin{align}
\label{eq:summation}
 S^{\rm conn}(t_s, \mathbf{q}) 
	& \stackrel{\hphantom{t_s\gg 1}}{=}  a  \sum_{t=a}^{t_s-a}
	G_A^{\mathrm{conn,eff}}(t, t_s,\mathbf{q}) \\ \notag
  & \stackrel{t_s\gg a}{=} b_0(Q^2) + t_s G_A^{\rm conn}(Q^2) 
  \, +\mathcal{O}(t_s e^{-\Delta t_s}) \, .
\end{align}
We fit simultaneously for the ranges in $Q^2$ and $t_s\geq t_{s,\mathrm{min}}$ for all the choices of minimum source-sink
separation $ t_{s,\mathrm{min}}$  using a $z$-expansion parameterization
\begin{align}
\label{eq:GA-zexp}
G^{\rm conn}_A(Q^2) &= \sum_{k=0}^{n} a_k z^{k}(Q^2) \, , \\
\quad z(Q^2) &= \frac{\sqrt{t_{\rm cut}+Q^2}-\sqrt{t_{\rm cut}}}{\sqrt{t_{\rm cut}+Q^2}+\sqrt{t_{\rm cut}}} \, ,
\end{align}
where we set $n=2$.
The fit parameters are therefore $b_0(Q^2)$ for each $Q^2$ and the three coefficients of the $z$-expansion $a_k$. 
We set  $t_{\rm cut} = (4M_{\pi})^2$ 
using the physical pion mass $M_\pi\equiv m_{\pi^0}^{\mathrm{phys}} = 134.977\,\text{MeV}$ across
all the ensembles in order to simplify the chiral extrapolation.
We note that this choice is common to both $A_\mu^8$ and $A_\mu^0$, while it differs from the one for $A_\mu^3$, where $t_{\rm cut} = (3M_{\pi})^2$.
This choice also defines the hadronic scheme adopted in this work.
Analogously to our previous determination, we perform the extrapolation using data up to $Q^2=0.7\,\text{GeV}^2$, after
which we observe a substantial increase in the statistical noise. The extracted connected contribution
is thus obtained in a fully consistent way with respect
to our previous analyses. In particular, we apply the same strategy as in Ref.~\cite{Barone:2025rye}
to deal with large and ill-conditioned covariance matrices, namely by applying
singular value decomposition (SVD) cuts to reduce the condition number of the matrices
for each ensemble.

\begin{figure}
\centering
\includegraphics[scale=0.28]{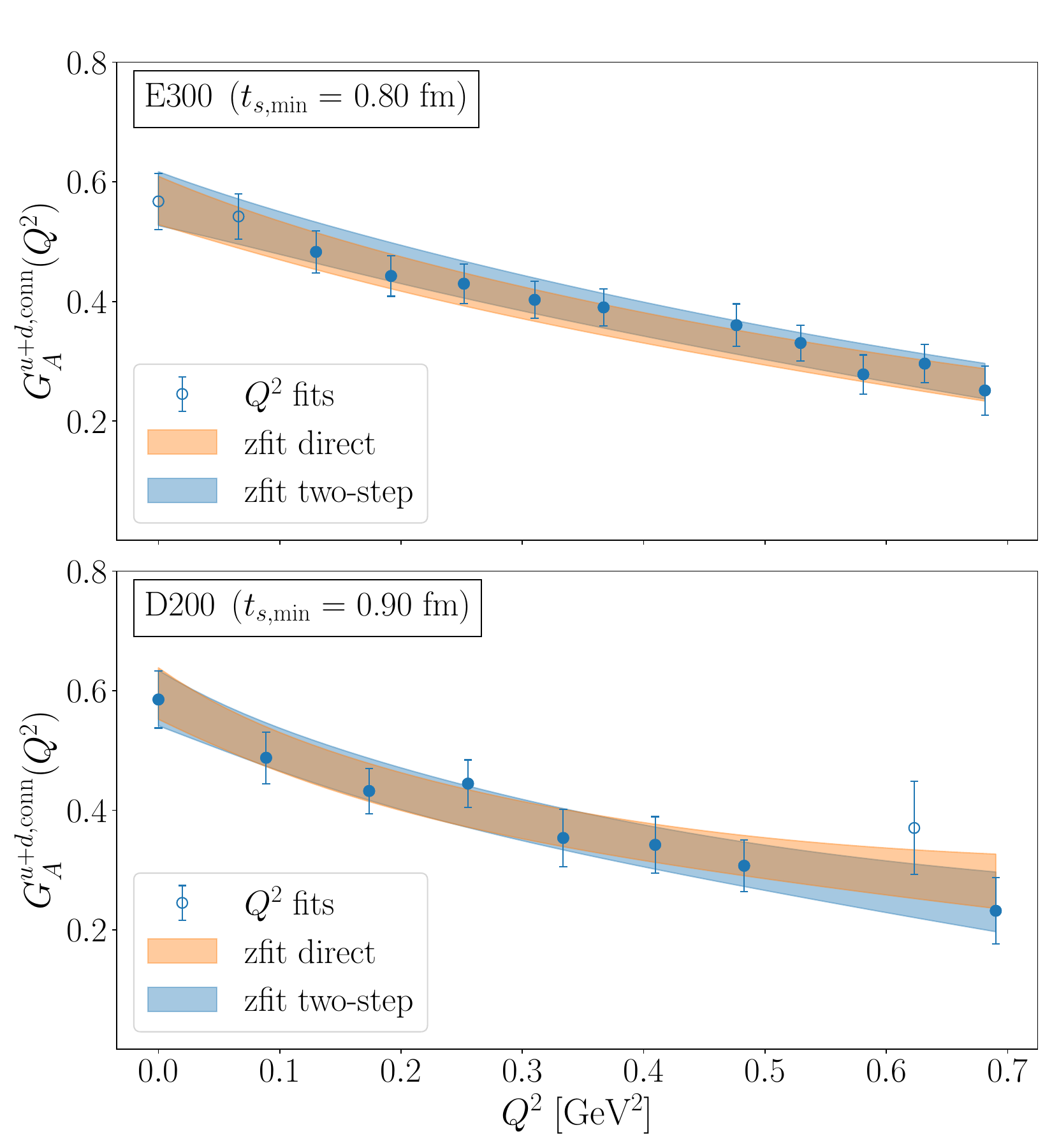}
\caption{Comparison of the direct $z$-fit and the two-step procedure for the 
bare connected contribution to the singlet form factor on the
ensembles E300  at $t_{s,\mathrm{min}}=0.80\,\text{fm}$ (top) and
D200  at $t_{s,\mathrm{min}}=0.90\,\text{fm}$ (bottom).
The empty points indicate a p-value smaller than $5\%$  or larger than $95\%$ for the first step of the two-step procedure, i.e. the linear fit to the
summation expression in~\eqref{eq:summation}, as exemplified in~\figref{fig:summation_example}.
}
\label{fig:plot_zfit_3ens}
\end{figure}

As an example,  we show in~\figref{fig:plot_zfit_3ens} the results from such a procedure for the connected contributions
of two of our more chiral ensembles, comparing it with the more conventional 
``two-step'' approach to confirm their stability and reliability in the presence
of a (regularized) large covariance matrix. Note that some of the points correspond
to a fit with a p-value smaller than 0.05 or larger than 0.95:
this is often due to strong correlations in the data corresponding to different $t_s$,
which lead to very small $\chi^2$, or small statistical fluctuations in some of these data.
We show an example of such a summation-method fit in Appendix.

We continue the practice employed in our previous work \cite{Djukanovic:2022wru,Djukanovic:2023beb,Barone:2025rye} and determine our
final results for the $z$-expansion coefficients through
a weighted average obtained by assigning the weights through a window function $W$ defined as
\begin{align}
\label{eq:window}
 W(t_{s,\mathrm{min}}; t_w^{\rm low}, t_w^{\rm up} )=\frac{1}{N_w}\Big[ &\tanh\left(\frac{t_{s,\mathrm{min}}-t_w^{\rm low}}{\Delta t_w} \right) \\ \notag
        & - \tanh\left(\frac{t_{s,\mathrm{min}}-t_w^{\rm up}}{\Delta t_w} \right) \Big] \, ,
\end{align}
where $N_w$ is a normalization factor and 
$t_w^{\rm low} = 0.75 \, \text{fm}$, $t_w^{\rm up} = 0.95\, \text{fm}$, $\Delta t_w = 0.1\, \text{fm}$
on each ensemble, in analogy to the window function chosen for the octet case~\cite{Barone:2025rye}.

\subsection{Disconnected Contribution}
\label{subsec:disc}

The disconnected contributions exhibit a much higher level of statistical noise compared to their connected counterparts.
The ratio in~\eqref{eq:ratio} leads thus to noisy plateaus with no detectable excited-state contamination 
within errors, as shown in~\figref{fig:disc_octet_VS_singlet} (bottom). For this reason, we choose to perform plateau fits to extract
the disconnected contribution $G^{\rm disc}_A(t_s,\bm{q})$.

Specifically, for each source-sink separation $t_s$ we fit 
the effective form factor $G_A^{\rm eff,disc}(t,t_s,\bm{q})$ 
for both light and strange contributions to a constant in $t$. To study the dependence on
the choice of fit range, we exclude at most two data points from
the outer edges of the interval.
We retain the fit that provides the
best quality as determined by the reduced $\chi^2$. The final result $G^{\rm disc}_A(Q^2)$ is then determined in a similar
way to the connected part through a weighted average over $t_s$
with the help of the window function $W(t_s;t_w^{\rm low}, t_w^{\rm up})$ in~\eqref{eq:window}, where in this case
we pick
$t_w^{\rm low} = 1.1 \, \text{fm}$, $t_w^{\rm up} = 1.3\, \text{fm}$ and $\Delta t_w = 0.1\, \text{fm}$
on each ensemble.

\begin{figure}
\centering
\includegraphics[scale=0.3]{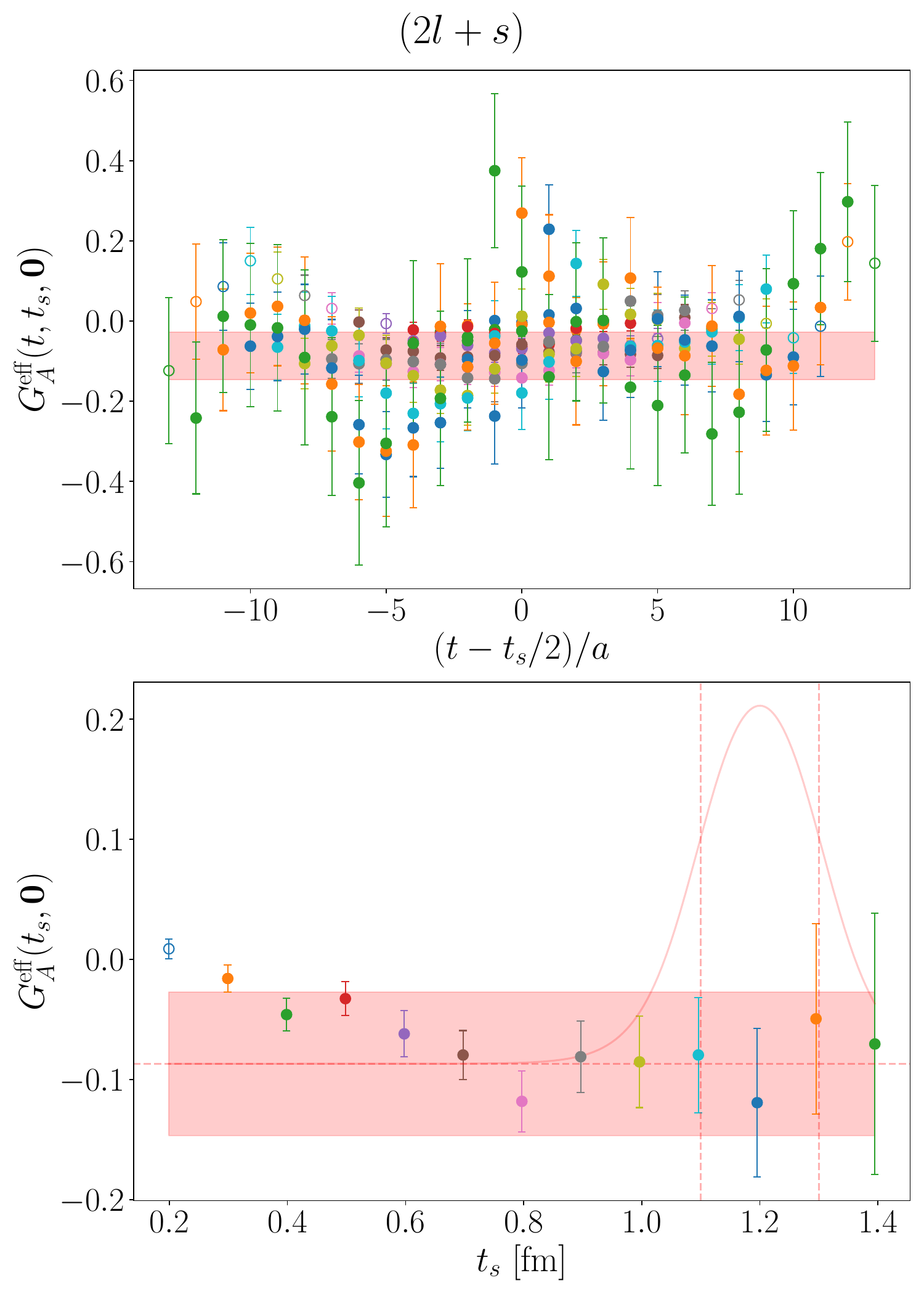} 
\caption{Example of the fitting procedure on the bare disconnected contribution $(2l+s)$
on E300 at vanishing momentum. The top panel shows the contribution to the effective form 
factor for the different source-sink separation $t_s$. The filled symbols indicate the
points that were used for the constant fits, whose results are shown in the bottom panel
together with the window average indicated by the red band. The red curve indicates
the window function used for the weighted average, and the vertical lines
correspond to $t_w^{\rm low}$ and $t_w^{\rm up}$.}
\label{fig:plateau_disc}
\end{figure}

The procedure is illustrated in~\figref{fig:plateau_disc} for the case of the bare $(2l+s)$ contribution at vanishing momentum on
the ensemble E300, highlighting the noise problem affecting the
disconnected contributions. In the lower panel we show the result for
the window average over different plateau fits, which is identified
with the region where the ratio exhibits no statistically significant
dependence on $t_s$, thereby resulting in a conservative estimate of
$G_A^{\rm disc}(Q^2)$.

\subsection{Combination of Contributions}

The total axial form factor is obtained by combining the connected and disconnected contributions.
In particular, after having obtained the contribution $G^{\rm disc}_A(Q^2)$ as in Sec.~\ref{subsec:disc},
we combine it with the result from the direct fit to the summation expression in~\eqref{eq:summation} 
as described in Sec.~\ref{subsec:connected}, 
\begin{align}
 G_A(Q^2) = G^{\rm conn}_A(Q^2) + G^{\rm disc}_A(Q^2) \, ,
\end{align}
i.e. we combine the $z$-expansion coefficients obtained from the different contributions.
The results are listed in Tabs.~\ref{tab:coeff_s} and~\ref{tab:coeff_singlet} in the Appendix.

\begin{figure}
\centering
\includegraphics[scale=0.3]{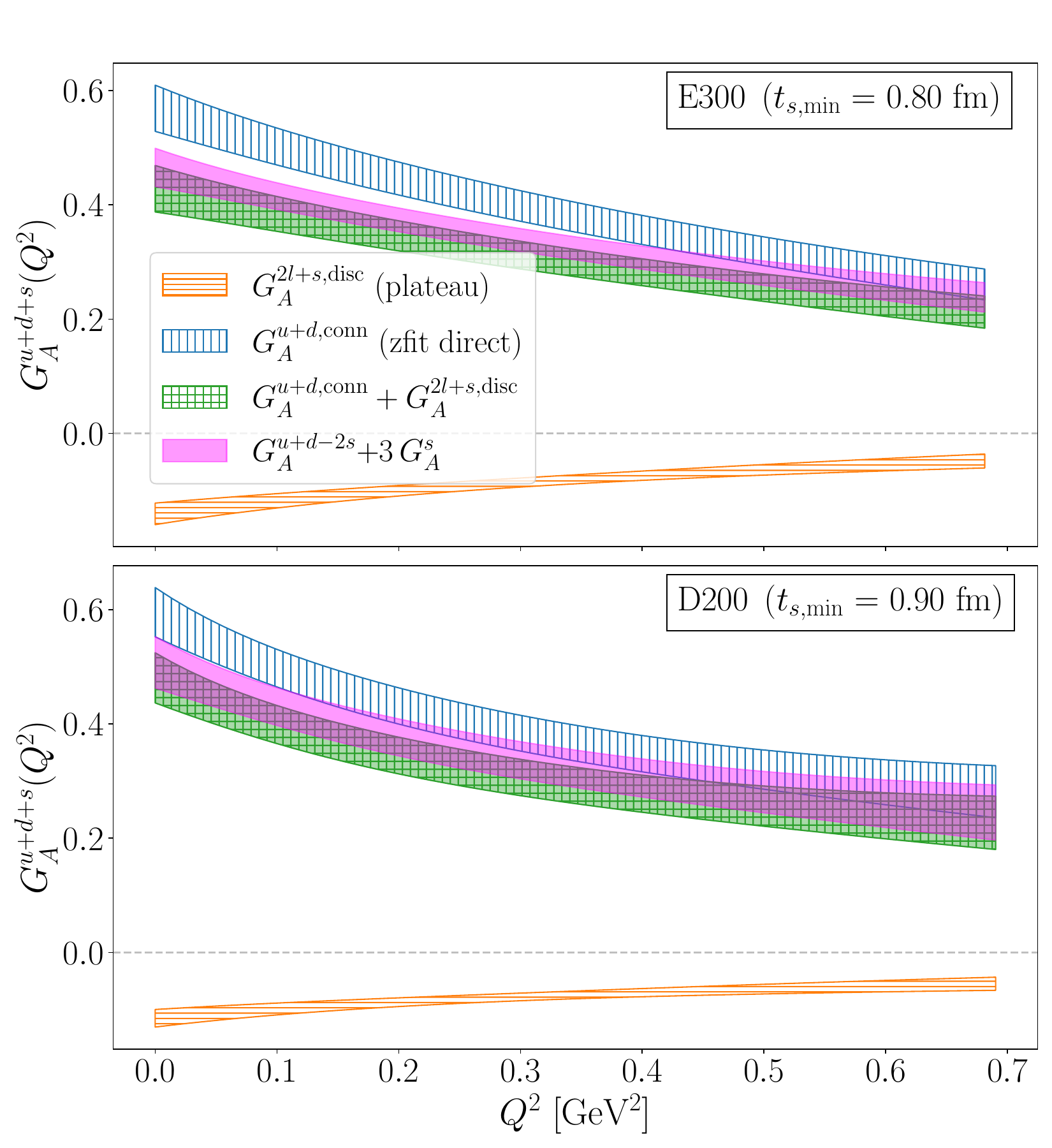}
\caption{The singlet axial form factor obtained with $z$-fit for E300 at $t_{s,\mathrm{min}}=0.80\,\text{fm}$ (top) 
and D200 at $t_{s,\mathrm{min}}=0.90\,\text{fm}$ (bottom).
We compare our chosen approach of adding the strange form factors $G_A^{s}$ to the isoscalar octet  $G_A^{u+d-2s}$ (magenta)
with the full result (green) obtained by treating the connected contribution (blue)
and the disconnected one (orange) separately,
as described in Secs.~\ref{subsec:connected} and \ref{subsec:disc}.
}
\label{fig:connVSdisc}
\end{figure}

In particular, the key analysis of this work focuses on the treatment of the disconnected contribution. First of all,
we determine the strange form factor $G_A^{s}(Q^2)$ in~\eqref{eq:AsR}, which is dominated by the disconnected contribution, since
the renormalization factor in~\eqref{eq:ZAs_conn} is numerically small.
Then, we determine the singlet by combining (on each ensemble) the results for the octet~\cite{Barone:2025rye}
and the strangeness, i.e. $G_A^{u+d+s}=G_A^{u+d-2s}+3G_A^{s}$. An example of this procedure is shown in~\figref{fig:connVSdisc},
where our chosen procedure (magenta) is compared to the sum (green) of individually determined connected and disconnected
contributions to the singlet (blue and orange, respectively).

\begin{figure*}[tp]
 \centering
 \includegraphics[scale=0.23]{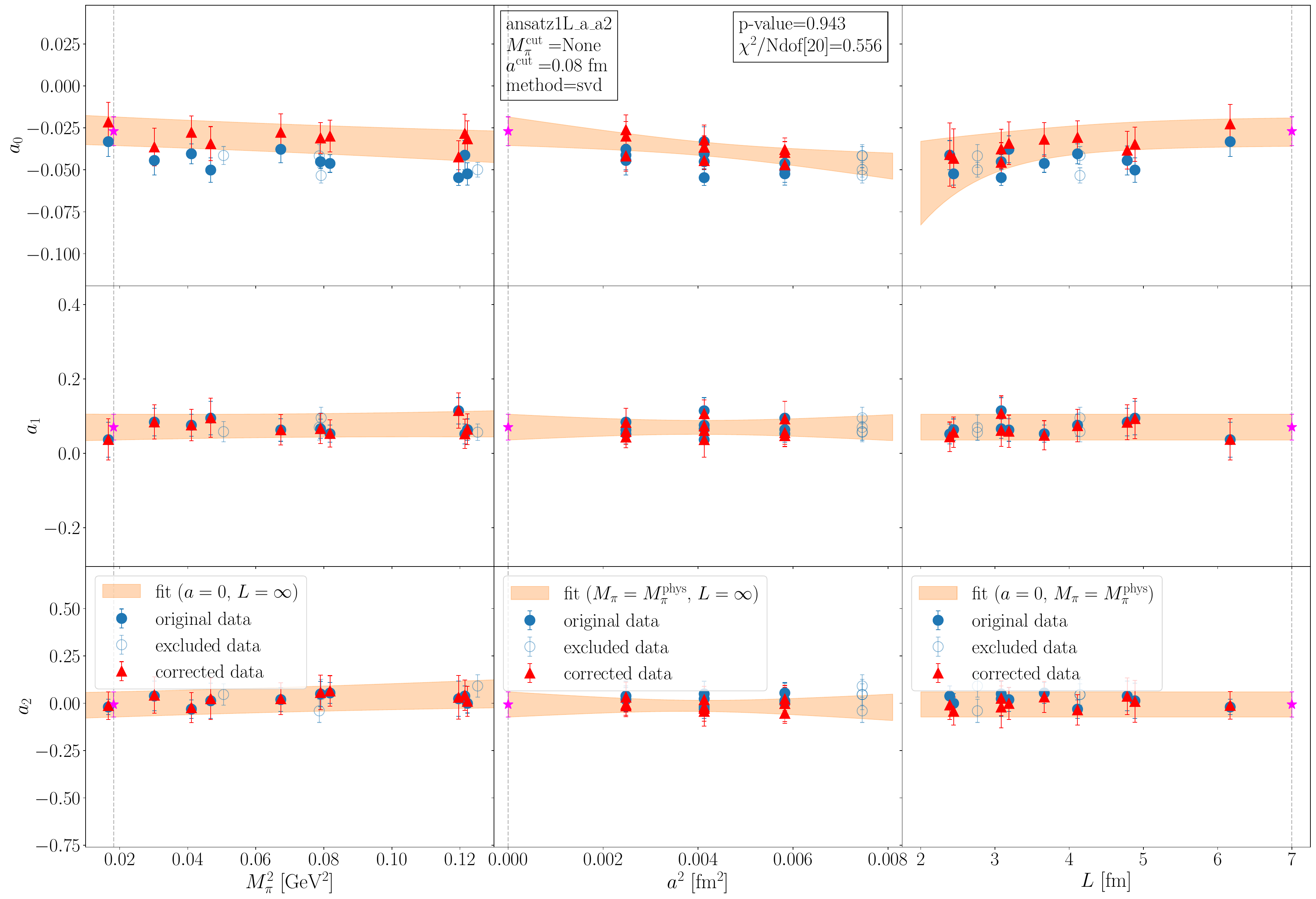}

\vspace{0.5cm}
\hspace{0.1cm}
\includegraphics[scale=0.23]{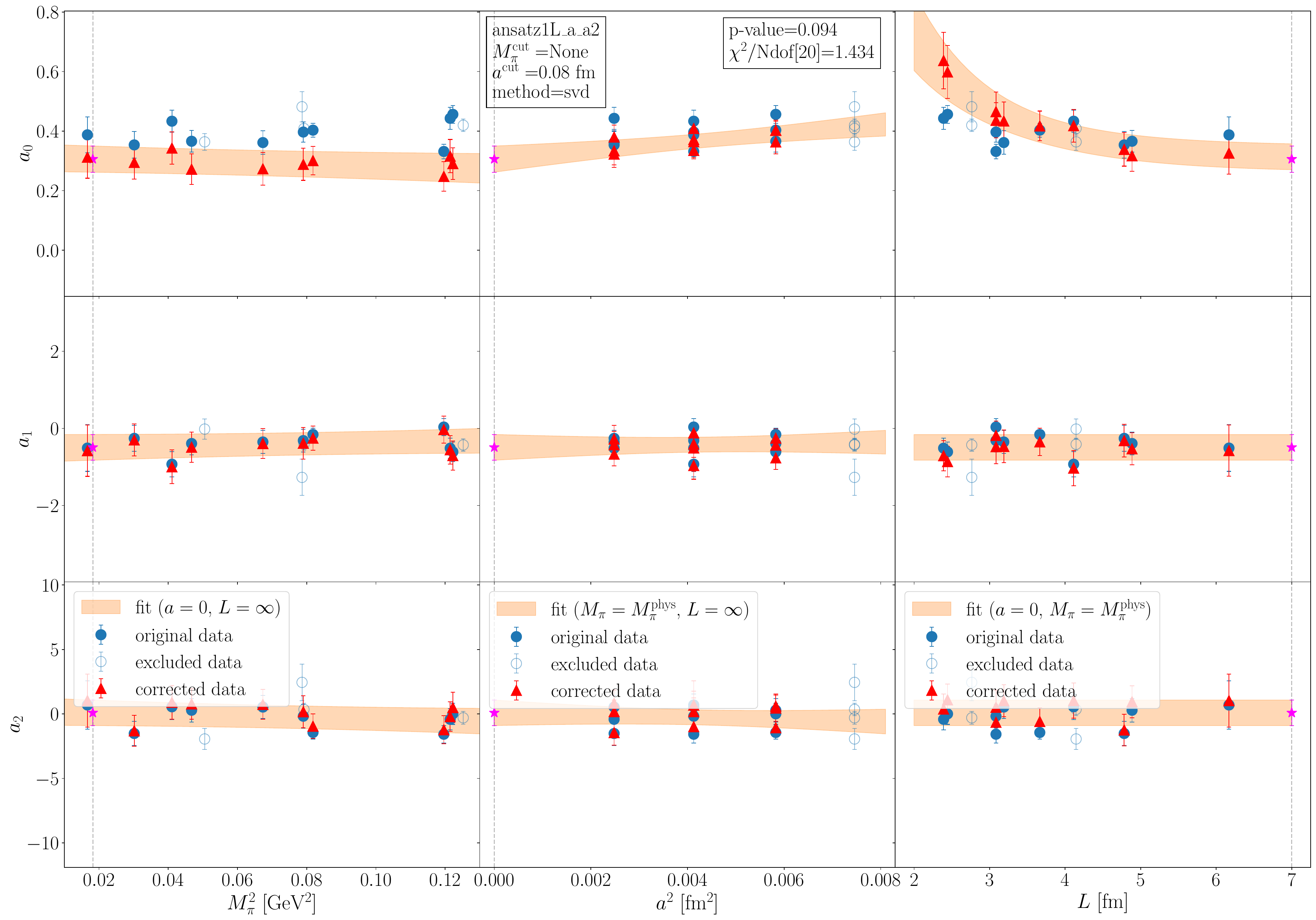}
\caption{Chiral continuum extrapolation for the strange (top) and singlet (bottom)
 with finite-volume term and with no cut on $M_{\pi}$ and without the coarsest lattice spacing. The results are plotted as a function of
 $M^2_{\pi}$ (left), $a^2$ (center) and $L$ (right). The blue circles indicate the original data, and the filled ones are the ones used after applying the cuts; the red triangles
 represent the corrected data with respect to the fit ansatz, as indicated in the legend for the orange bands.}
 \label{fig:cc_extr}
\end{figure*}

\subsection{Physical-point extrapolation and model average}
\label{subsec:extrap}

In order to obtain the final result, we
extrapolate the final coefficients of the $z$-expansions for both singlet and strange form factors to the chiral and continuum
limit using an ansatz linear in $M^2_\pi$ and $a^2$ for all the coefficients $a_i$, i.e.
\begin{align}
 a_i = d_{i,0} + d_{i,\pi} M^2_\pi + d^{(1)}_{i,a} \frac{a}{\sqrt{t_0}} + d^{(2)}_{i,a} \frac{a^2}{t_0} \, ,
\end{align}
including a linear term in $a$ to account for effects due to the mass-independent determination
of the renormalization constant $Z_A^{00}$. In practice, since we expect our data to be dominated by $\mathcal{O}(a^2)$ effects,
we impose a prior $d^{(1)}_{i,a}\sim \mathcal{N}(0, 0.1\cdot |{d^{(2)}_{i,a}}|)$, where $|{d^{(2)}_{i,a}}|$ refers to the absolute value
of the coefficient $d^{(2)}_{i,a}$ determined by a fit with $d^{(1)}_{i,a}=0$.
We include the term~\cite{Beane:2004rf}
\begin{align}
 \frac{M_\pi^2}{\sqrt{M_\pi L}}e^{-M_\pi L}  
\end{align}
in the axial charge $a_0$ to account for possible finite-volume effects.\footnote{We do not include any higher order terms~\cite{Hall:2025ytt} as we are not sensitive to them with our current statistics.}
We perform multiple fits with cuts in the pion mass, $M^{\rm cut}_{\pi}[\text{MeV}]=\{300, 285, 265\}$, and by removing the coarsest lattice spacing, preserving the correlations
among the three coefficients on each ensemble.

This ansatz appears to be sufficient to describe the data well thanks to their flat behavior in the variables $M^2_\pi$, $a^2$ and the
lattice spatial extent $L$. We show an example of the chiral-continuum extrapolation in~\figref{fig:cc_extr}.

To finalize the results, we employ a model average~\cite{Jay:2020jkz} procedure based on
the Akaike Information Criterion (AIC)~\cite{Akaike}. Each fit model $k$
is assigned a weight
\begin{align}
 w^{\rm AIC}_k \propto 
 e^{-\frac{1}{2}(\chi^2_k +2n_{{\rm par}, k} -n_{{\rm data}, k} )}  \, ,
\end{align}
with $n_{{\rm par}, k}$ being the number of parameters and $n_{{\rm data}, k}$ the number of data points entering the fit.
As discussed in~\cite{Barone:2025rye}, given the gaussianity of the weighted joint distribution
built out of the above values $w^{\rm AIC}_k$, we perform a weighted average of the different models,
thus accounting for both statistical and systematic uncertainties.

\section{Results\label{sec:results}}

\begin{figure}[t]
\centering
\includegraphics[scale=0.32]{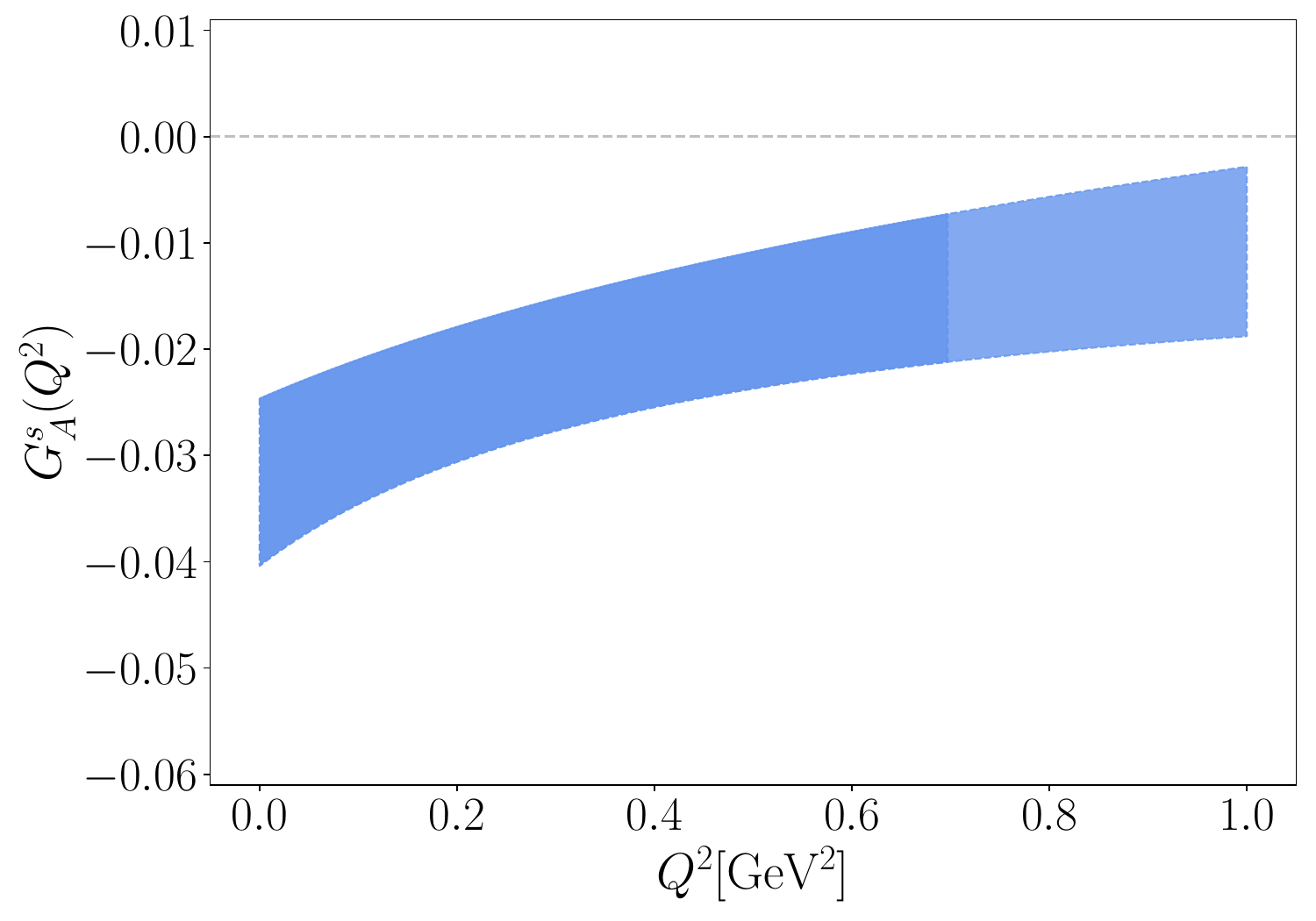}

\vspace{0.2cm}
\hspace{0.25cm}
\includegraphics[scale=0.305]{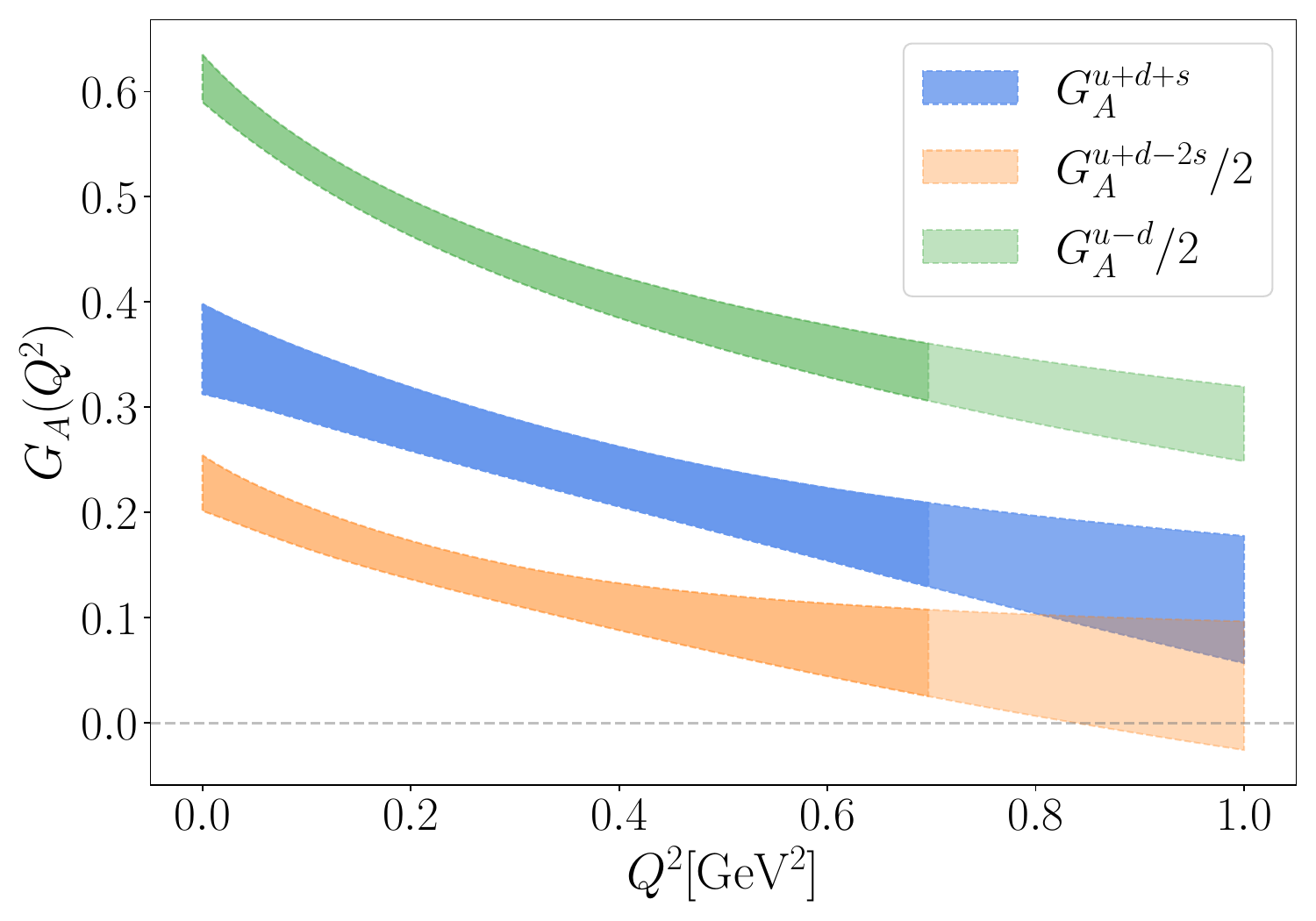}
 \caption{Final result for the strange (top) and singlet (bottom) form factors.
 The data are used up to $Q^2=0.7\, \text{GeV}^2$, and the lighter-shaded bands represent extrapolations based
   on the respective parameterizations. In the lower panel, we also plot the rescaled isovector and isoscalar octet
   form factors $G_A^{u-d}/2$~\cite{Djukanovic:2022wru} and $G_A^{u+d-2s}/2$~\cite{Barone:2025rye} for comparison.}
 \label{fig:GA_final}
\end{figure}

\begin{figure}[t!]
\centering
\includegraphics[scale=0.45]{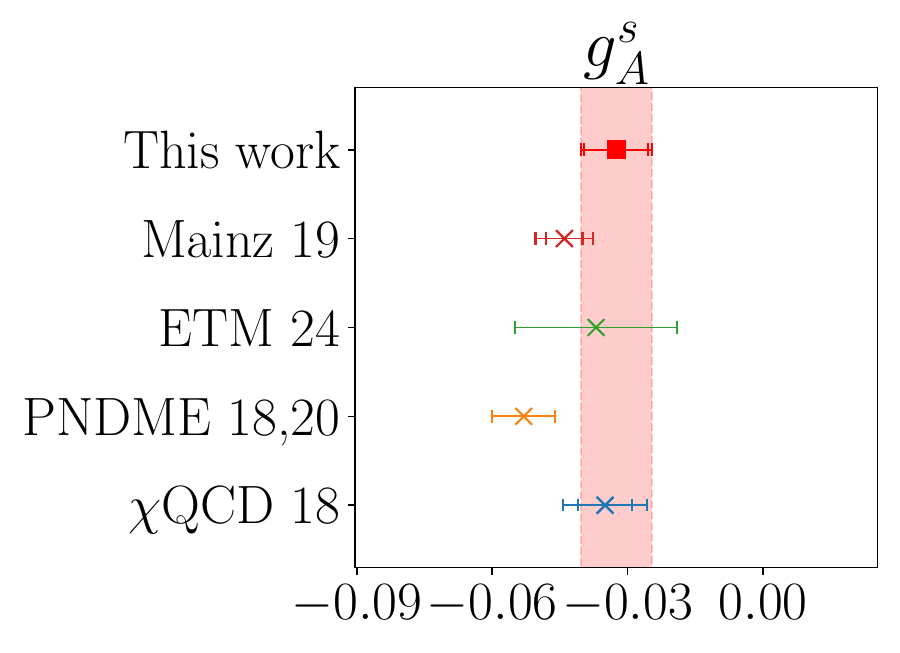}
\caption{Our result for the axial charge $g_A^s$ (red square) compared to other collaborations. The red cross
 shows our previous result~\cite{Djukanovic:2019gvi}.
The green point shows the result from ETM~\cite{Alexandrou:2024ozj},
the orange from PNDME~\cite{Lin:2018obj,Park:2020axe}, and the blue from $\chi$QCD~\cite{Liang:2018pis}.}
\label{fig:gAs}
\end{figure}

Our results for the coefficients of the $z$-expansion of
the strange nucleon axial form factor in the continuum and at the physical pion mass are
\begin{align}
\begin{split}
a^s_0 &= -0.0325 \pm 0.0071 \, \text{(stat)} \pm  0.0033 \,\text{(sys)} \,,\\ 
a^s_1 &= +0.0651 \pm 0.0310 \, \text{(stat)} \pm  0.0077 \,\text{(sys)} \,,\\ 
a^s_2 &= -0.0116 \pm 0.0592 \, \text{(stat)} \pm  0.0048 \,\text{(sys)} \,,
\end{split}
\end{align}
with correlation matrix
\begin{align}
C^s=
\begin{pmatrix}
1.00000 & -0.57655 & 0.11734 \\ 
-0.57655 & 1.00000 & -0.56446 \\ 
0.11734 & -0.56446 & 1.00000 
\end{pmatrix}\,.
\end{align}
The result is plotted in~\figref{fig:GA_final} (top) and extrapolated up to $Q^2=1\,\text{GeV}^2$ (as indicated by the shaded areas).
The charge is then given by
\begin{align}
 g_A^{s}&=  -0.0325(78) \,,
\end{align}
and it is reported in comparison to other recent results in~\figref{fig:gAs}.
We also note that the FLAG review 2024~\cite{FlavourLatticeAveragingGroupFLAG:2024oxs} quotes averages for
$N_f=2+1$ and $N_f=2+1+1$ values for the strange axial charge, which are
identified with the single $\chi$QCD 18~\cite{Liang:2018pis} and PNDME 18~\cite{Lin:2018obj} calculations, respectively.

The singlet case has been obtained through a combination of the octet channel and the strange component, taking advantage of
the better signal quality of the octet, which allows us to use the summation method on the whole dataset. The final result is given by
\begin{align}
\begin{split}
a^{u+d+s}_0 &= +0.355 \pm 0.041 \, \text{(stat)} \pm  0.013 \,\text{(sys)} \,,\\ 
a^{u+d+s}_1 &= -0.423 \pm 0.322 \, \text{(stat)} \pm  0.064 \,\text{(sys)} \,,\\ 
a^{u+d+s}_2 &= -0.691 \pm 0.947 \, \text{(stat)} \pm  0.448 \,\text{(sys)} \,,\\ 
\end{split}
\end{align}
with correlation matrix
\begin{align}
C^{u+d+s} = 
\begin{pmatrix}
1.00000 & -0.57561 & 0.22890 \\ 
-0.57561 & 1.00000 & -0.85035 \\ 
0.22890 & -0.85035 & 1.00000 
\end{pmatrix} \, .
\end{align}
Our result for the singlet axial charge is 
(in the $\overline{\text{MS}}$ scheme at a renormalization scale of 2 GeV) 
\begin{align}
g_A^{u+d+s}= 0.355(43),
\end{align}
and the full form factor is plotted in ~\figref{fig:GA_final} (bottom).
Therefore, our result suggests that the intrinsic quark spin contributes roughly $35\%$ to the proton spin,
while the remaining fraction must be supplied by gluon and orbital angular momentum contributions.
Historically, the polarized deep-inelastic scattering experiment EMC~\cite{EuropeanMuon:1987isl,EuropeanMuon:1989yki} already found a value on the order of 30\%.
Our result is fully consistent with the determination based on the more recent deep-inelastic scattering data of the COMPASS experiment~\cite{COMPASS:2006mhr}
supplemented by an estimate of the octet axial charge. Subtracting from $g_A^{u+d+s}$ the SU(3)$_f$-symmetry estimate of $g_A^{u+d-2s}$ based on hyperon beta decays is, however, too crude to reliably determine the (small) strangeness contribution to the nucleon spin~\cite{Aidala:2012mv}. 

\section{Conclusions\label{sec:conclusions}}

This work provides the first lattice QCD determination with a full error budget of the
singlet axial form factor  $G_A^{u+d+s}(Q^2)$ of the nucleon in a large $Q^2$ range, together with the
strange $G_A^{s}(Q^2)$ contribution.
The calculation completes our program of studies of the nucleon axial form factor across flavor channels,
following our previous determinations of the isovector~\cite{Djukanovic:2022wru} and isoscalar octet~\cite{Barone:2025rye}
combinations performed within the same lattice setup.

The singlet and the strange channels introduce qualitatively new challenges compared to the
non-singlet cases, most notably the presence of  noisy disconnected quark-loop
contributions. In this work, the connected contribution is analyzed using the
summation method to suppress excited-state contamination, while the disconnected
contribution is determined using plateau fits to the effective form factor. 

The momentum dependence of the form factor is parameterized
using the $z$-expansion, and the results for its coefficients 
are extrapolated to the physical point through a linear ansatz in $M_\pi^2$ and $a^2$ (plus a small parameterization for possible $O(a)$ effects).
Our final result is obtained through a model average inspired by the AIC and includes a complete error budget that incorporates both
statistical and systematic uncertainties.

Together with our previous results, this work provides a comprehensive lattice QCD determination of
the nucleon axial structure across flavor channels within a unified
computational framework. 
Future extensions of this program will include a full flavor decomposition
through a unified treatment of the three channels with the use of $\bm{p}'\neq \bm{0}$ frames.
In this way, the charges of the single-flavor components can be extracted with the best possible statistics,
providing insight into the composition of the proton spin.

\section*{Acknowledgments}
This work was supported
by the Deutsche
  Forschungsgemeinschaft (DFG) through the Collaborative Research
  Center  1660 ``Hadrons and Nuclei as Discovery Tools'',
  under grant HI~2048/1-3 (Project No.\ 399400745) and in the Cluster
  of Excellence “Precision Physics, Fundamental Interactions and
  Structure of Matter” (PRISMA++ EXC 2118/1) funded by the DFG within
  the German Excellence strategy (Project ID 39083149).
  Calculations for this project were partly performed on the HPC
  clusters ``Clover'' and ``HIMster2'' at the Helmholtz Institute Mainz,
  and ``Mogon 2'' at Johannes Gutenberg-Universit\"at Mainz.
  The authors gratefully acknowledge the Gauss Centre for Supercomputing e.V. (www.gauss-centre.eu) 
  for funding this project by providing computing time on the GCS Supercomputer systems JUQUEEN and JUWELS at J\"ulich Supercomputing Centre (JSC) 
  via grants HMZ21, HMZ23 and HMZ36 (the latter through the John von Neumann Institute for Computing (NIC)), as well as on the GCS Supercomputer HAZELHEN at
   H\"ochstleistungsrechenzentrum Stuttgart (www.hlrs.de) under project GCS-HQCD.
  
Our programs use the QDP++ library~\cite{Edwards:2004sx} and deflated SAP+GCR
solver from the openQCD package~\cite{Luscher:2012av}, while the contractions
have been explicitly checked using~\cite{Djukanovic:2016spv}. We are grateful to
our colleagues in the CLS initiative for sharing the gauge field configurations
on which this work is based.

\cleardoublepage
\section*{\label{sec:appendix}Appendix}
\subsection{Renormalization}

Our calculation of the renormalization factors is described in detail in
Ref.~\cite{Harris:2019bih}; here we summarize the pertinent formulae and
concentrate on the
results for the singlet case, for which we follow the preliminary work presented in~\cite{wilhelm_2020}. 
We list the set of $N_f=3$ ensembles employed in~\tabref{tab:nprens}.

We work in the RI$'$-MOM scheme
\cite{Martinelli:1994ty}, where the renormalization condition is specified by
matching the renormalized matrix element of an operator $O_\Gamma$ between two
quark states of momentum $p$ at vanishing momentum transfer to its tree-level
value \cite{Gockeler:1998ye} 
\begin{align}
Z_O\langle p |O_\Gamma | p \rangle |_{p^2=\mu^2} &=  \langle p | O_\Gamma |p
\rangle_{\text{tree}}|_{p^2=\mu^2}.
\end{align} 
We consider the flavour diagonal currents for a generic $\Gamma$, i.e.
\begin{align}
O_\Gamma^a(x) = \bar \psi (x) \Gamma \lambda ^a \psi(x)  \,.
\end{align}
The matrix element is related to the amputated Green functions, which read
\begin{align}
\Lambda_{O_\Gamma^a}^f(p) &= S^{-1}_f(p) G_{O_\Gamma^a}^f(p) S^{-1}_f(p),
\nonumber\\
G_{O_\Gamma^a}^f(p)        &= \frac{1}{V}\sum\limits_{x,y,z} e^{-ip(x-y) } 
\langle f(x) O_\Gamma^a(z) \bar{f}(y)\rangle ,
\label{eq_amputated_green_functions}
\end{align}
where $f$ denotes the quark field with a given flavour, and $S_f(p)$ is the quark propagator
\begin{align}
 S(p) = \frac{1}{V}\sum_{x,y} e^{-ip(x-y)} \langle f(x) \bar{f}(y) \rangle \, .
\end{align}
For the renormalization condition we use 
\begin{align}
Z_A&=\frac{12 Z_q}{\frac{1}{3} \sum\limits_{\mu\nu} 
\Bigl( \delta_{\mu\nu} - \frac{p_\mu p_\nu}{p^2} \Bigr) \text{tr}
\left. \Bigl[\Lambda_{A_\mu}(p) \Lambda^{-1,\text{tree}}_{A_\nu}(p) \Bigr]\right|_{p^2=\mu^2}
}
\end{align}
which is compatible with the axial Ward identity \cite{Green:2017keo}.
The renormalization constants for the singlet case are then obtained from the
generalized amputated Green functions in~\eqref{eq_amputated_green_functions}, which renormalize as~\cite{Martinelli:1994ty,Green:2017keo}
\begin{align}
\Lambda_{O_\Gamma^a}^f(p)_R&= \frac{Z^{ab}_\Gamma}{Z_q}
\Lambda_{O_\Gamma^b}^f(p).
\label{eq_op_ren}
\end{align}
$Z_q$ denotes the quark normalization constant, which
is determined by matching the renormalized quark propagator to the free massless
lattice propagator, i.e. 
\begin{align}
Z_q &=  \frac{1}{12} \text{tr} \Bigl [ S^{-1} (p ) S_{\text{free}}(p) \Bigr
]|_{p^2=\mu^2} \, ,
\end{align}
with
\begin{align}
S_{\text{free}}(p)&= \frac{- i a \sum\limits_\mu \gamma_\mu \sin (a
p_\mu)}{\sum\limits_\mu \sin^2 (ap_\mu)}.
\end{align}

\begin{table}
 \centering
 \begin{tabular}{llllllc}
  \hline\hline
  ID       & $\beta$ & $a/\mathrm{fm}$ & $T/a$ & $L/a$ & $\kappa$ & $M_\pi/\mathrm{MeV}$ \\
  \hline
  rqcd.019 & 3.40 & 0.086 & 32 & 32 & 0.1366     & 600 \\
  rqcd.016 & 3.40 & 0.086 & 32 & 32 & 0.13675962 & 420 \\
  rqcd.021 & 3.40 & 0.086 & 32 & 32 & 0.136813   & 340 \\
  rqcd.017 & 3.40 & 0.086 & 32 & 32 & 0.136865   & 230 \\
  \hline
  rqcd.029 & 3.46 & 0.076 & 64 & 32 & 0.1366     & 700 \\
  rqcd.030 & 3.46 & 0.076 & 64 & 32 & 0.1369587  & 320 \\
  X450     & 3.46 & 0.076 & 64 & 48 & 0.136994   & 250 \\
  \hline
  B250     & 3.55 & 0.064 & 64 & 32 & 0.1367     & 710 \\
  B251     & 3.55 & 0.064 & 64 & 32 & 0.137      & 420 \\
  X250     & 3.55 & 0.064 & 64 & 48 & 0.13705    & 350 \\
  X251     & 3.55 & 0.064 & 64 & 48 & 0.13710    & 270 \\
  \hline\hline
 \end{tabular}
 \caption{\label{tab:nprens}The $N_{\rm f}=3$ flavor ensembles with periodic boundary conditions used to determine the renormalization constants.}
\end{table}

For flavor  diagonal currents and degenerate quark masses the renormalization
matrix is diagonal, i.e.
\begin{align}
Z^{ab}_\Gamma=
\begin{pmatrix}
Z^{33}_\Gamma & 0 & 0 \\
0 & Z^{88} _\Gamma &0 \\
0 & 0 & Z^{00}_\Gamma
\end{pmatrix}.
\end{align}
We note that $Z_\Gamma^{33}=Z_\Gamma^{88}$, and $Z_\Gamma^{88}$ is identical to
the isovector result.  Finally, the renormalization conditions read
\begin{align}
(Z_\Gamma^{-1})^{da}&=\sum\limits_{f=u,d,s} \text{diag}(\lambda^a)
\frac{1}{12  Z_q} \text{tr} \Bigl[
\Lambda^f_{O_{\Gamma^d}}(p) \Gamma^{-1} \Bigr]_{p^2=\mu^2} .
\end{align}
For the  flavor-diagonal currents  we obtain
\begin{align}
(Z_\Gamma^{-1})^{00}(p)&= \frac{1}{12 Z_q} \Bigl (\Sigma_{\text{con}}^\Gamma(p)
+ 3 \Sigma_{\text{dis}}^\Gamma(p) \Bigr),\\
(Z_\Gamma^{-1})^{88}(p)&=\frac{1}{12 Z_q} \Sigma_{\text{con}}^\Gamma(p),
\end{align}
where the connected and disconnected contributions of a single quark flavor read
\begin{align}
\begin{split}
\Sigma_{\text{con}}^\Gamma(p)
= \text{tr}\!\left[
S^{-1}(p)\,
\mathcal{O}^{\rm conn}_\Gamma(p)\,
S^{-1}(p)\,
\Gamma^{-1}
\right] \, , \\
\mathcal{O}^{\rm conn}_\Gamma(p)
= \Biggl\langle 
\frac{1}{V}\sum_z 
\gamma_5 \mathcal{S}(z,p)^\dagger \gamma_5 \Gamma
\mathcal{S}(z,p)
\Biggr\rangle \,,
\end{split}
\end{align}
for the connected and
\begin{align}
\begin{split}
\Sigma_{\mathrm{dis}}^{\Gamma}(p)
&=
-\,\operatorname{tr}\!\left[
S^{-1}(p)\,
\mathcal{O}_{\Gamma}^{\mathrm{dis}}(p)\,
S^{-1}(p)\,
\Gamma^{-1}
\right]\, , \\
\mathcal{O}_{\Gamma}^{\mathrm{dis}}(p)
&=
\Biggl\langle 
\mathcal{S}(p)\,
\frac{1}{V}\sum_z
\operatorname{tr}\!\left[
\mathcal{S}(z,z)\Gamma
\right]
\Biggr\rangle \,,
\end{split} 
\end{align}
for the disconnected. $\mathcal{S}$ refers to the estimate of the
propagator on each configuration, i.e. before taking the gauge average.
The traces are taken over spin and color indices, and the brackets denote the gauge
average. We obtain the renormalization constant for each value of the lattice
spacing at different quark masses.
We extrapolate the
renormalization constants to the massless limit, 
accounting for systematics by performing different chiral extrapolations, i.e. with or without a $M_\pi^4$-term and 
a term proportional to $e^{-M_\pi L}$, as well as different fit intervals in the scale $\mu$.
We then subtract lattice
artifacts, c.f. \cite{Harris:2019bih}, which to first order in the coupling coincide with the
isovector case.
As in \cite{Harris:2019bih} we explore several alternatives for defining the
coupling used for the subtraction, where the quoted values in~\tabref{tab_ren_consts}
are based on the BLM coupling of \cite{Brodsky:1982gc,Lepage:1992xa}. Finally we transform the renormalization constants to the RGI
scheme. For the singlet case we implement the one-loop anomalous dimensions
from Ref.~\cite{Larin:1993tq}, in the conversion factor $\Delta Z^{\overline{\text{MS}}}$ between the $\overline{\text{MS}}$ to the RGI
scheme. The 
renormalization constants are determined from a simultaneous fit to all lattice
data  using
\begin{align}
Z^{\text{RGI}}_{\text{sub}}(a,\mu)&=
Z^{\text{RGI}} (\beta) \Bigl \{ 1+ c_1 g^{\overline{\text{MS}}}(\mu)^4 \Bigr\}
\nonumber\\& +
c_2(\beta) \bigl(a\mu\bigr)^2 \Delta Z^{\overline{\text{MS}}}(\mu)
Z^{\overline{\text{MS}}}_{\text{RI}'-\text{MOM}}(\mu).
\label{RGI_eq}
\end{align}
We note that the fit constant $c_1$ is independent of $\beta$ and parametrizes
the neglected higher-order contributions in the conversion factors.
The central values of these fits are summarized in Tab.~\ref{tab_ren_consts},
where the value for $\beta=3.7$ is the result based on an extrapolation of the
other lattice spacings.
Indeed, employing periodic boundary conditions is essential for the
RI-MOM scheme, which is impossible at our finest lattice spacing,
where open boundary conditions are used to mitigate topological
freezing. 
Thus, in the absence of a viable
ensemble at our finest lattice spacing, we extrapolate the renormalization
constant using the result from the three other lattice spacings.

\begin{table}[t]
\begin{ruledtabular}
\begin{tabular}{c|ccc|c}
$ \beta$ &  $3.4$ & $3.46$  & $3.55$ & $3.7$ \\ 
\hline 
$Z_A^{00}$ & $0.721(6)(18)$ & $0.729(5)(21)$ & $0.743(5)(16)$ & $0.77(4)$ 
\end{tabular} 
\end{ruledtabular}
\caption{Renormalization constants for the axial singlet current on CLS $N_f=3$
ensembles in the RGI scheme. Note that the value for $\beta=3.7$ is the result of a linear
extrapolation based on the three coarser lattice spacings.}
\label{tab_ren_consts}
\end{table}

The singlet current renormalization is scheme dependent. In Table
\ref{tab_ren_consts} we quote the RGI value obtained from Eq.~(\ref{RGI_eq}).
For the final results we transcribe our results to the $\overline{\text{MS}}$ scheme
at $\mu=2\text{GeV}$. For the conversion from the RGI values we use the
anomalous dimensions of the singlet current from Ref.~\cite{Larin:1993tq}
\begin{align}
    \gamma(a_s)^{\text{singlet}}&= a_s^2 (-6 C_f n_f)+a_s^3 (-\frac{142}{3}C_F C_A n_f + \frac{4}{3} C_F n_f^2 \nonumber\\
    &+ 18 C_F^2 n_f),\nonumber\\
    &=a_s^2 \gamma_1  +a_s^3 \gamma_2,\label{eq_anomalous_dims_singlet}
\end{align}
where $a_s= \alpha_s/(4 \pi)$, $n_f=3$, $C_A=3$ and $C_F=4/3$. For the running of the strong coupling we use \cite{Larin:1993tp}
\begin{align}
   \beta_0 &= 11 -\frac{2}{3}n_f,\nonumber\\
    \beta_1 &= 102 - \frac{38}{3} n_f,\nonumber\\
    \beta_2 &= \frac{2857}{2} - \frac{5033}{18}n_f +\frac{325}{54} n_f^2,\nonumber\\
    \beta_3 &= \frac{149753}{6}+3564 \zeta_3 -\left(\frac{1078361}{162}+\frac{6508}{27}\zeta_3\right) n_f\nonumber\\
    &+
    \left(
        \frac{50065}{162}+\frac{6472}{81}\zeta_3
    \right)n_f^2+\frac{1093}{729}n_f^3
    \\
    a_s(\mu)&=\frac{1}{\beta_0 L }\left(
        1-\frac{\beta_1}{\beta_0^2}\frac{\ln L }{L}+\frac{\beta_1^2\ln^2 L
-\beta_1^2 \ln L - \beta_1^2+\beta_0\beta_2}{\beta_0^4 L^2}\right),
        \intertext{with}
        L &=\ln \frac{\mu^2}{\Lambda^2}.
\end{align}
The conversion of the RGI values to the $\overline{\text{MS}}$ values proceeds via
\begin{align}
Z_A^{\overline{\text{MS}}} &= \frac{Z_A^{\text{RGI}}}{\Delta Z^{\overline{\text{MS}}}(\mu)},\nonumber\\
\Delta Z^{\overline{\text{MS}}}(\mu) &= (2\beta_0 a_s)^{-\frac{\gamma_0}{2\beta_0}} \left( 1 + c_1 a_s + c_2 a_s^2 + \mathcal{O}(a_s^3) \right) \\
    c_1 &= \frac{\beta_1 \gamma_0 - \beta_0 \gamma_1}{2\beta_0^2} \\[10pt]
    c_2 &= \frac{1}{8\beta_0^4}\Bigl[\beta_1^2 \gamma_0^2 - 2\beta_0 \beta_1
    \gamma_0 (\beta_1 + \gamma_1) \nonumber\\
    &+ \beta_0^2 \left[ 2\beta_2 \gamma_0 +
    \gamma_1 (2\beta_1 + \gamma_1) \right] - 2\beta_0^3 \gamma_2\Bigr].
\end{align}
We note that $\gamma_0=0$ for the anomalous dimensions of the singlet current of
Eq.~(\ref{eq_anomalous_dims_singlet}). To evaluate the
strong coupling $a_s$ we use
$\Lambda_{\text{QCD}}=  344.4(8.7) \text{MeV}$ \cite{DallaBrida:2026kuo} as input. 

\cleardoublepage
\onecolumngrid
\subsection{Form factor data}

\begin{figure}[h]
\centering
\includegraphics[scale=0.3]{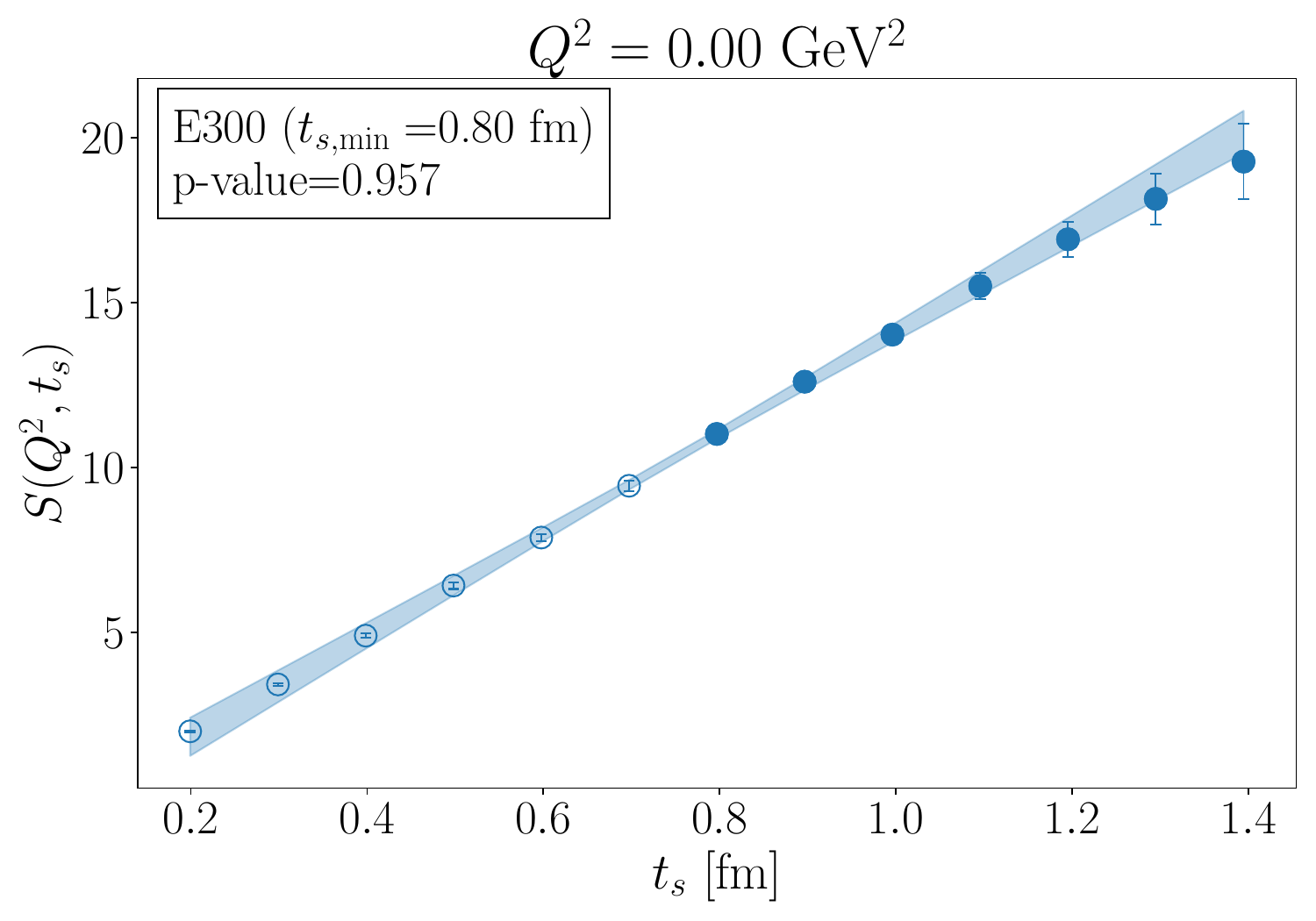}
\includegraphics[scale=0.3]{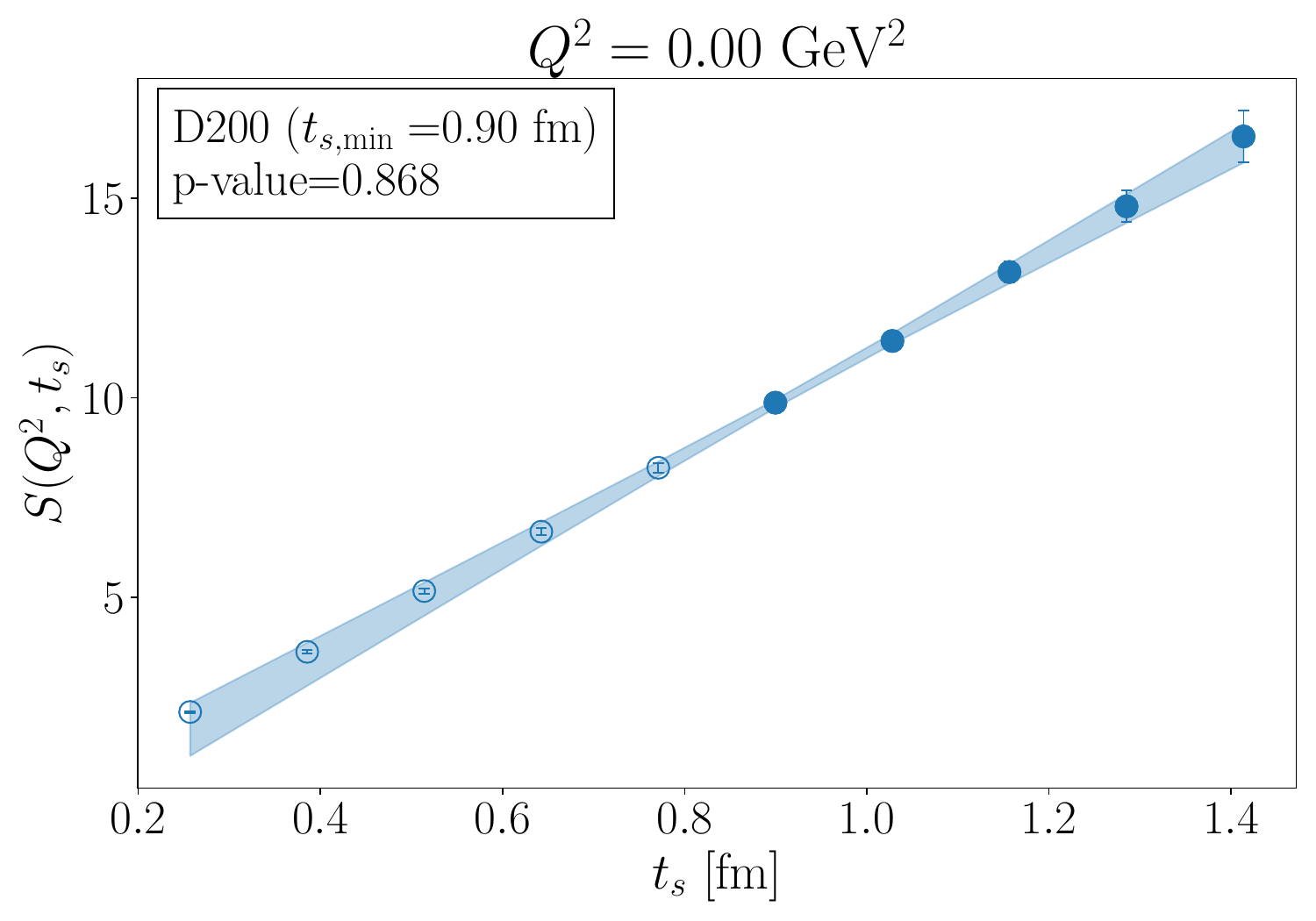}
\caption{Linear fit to the summation expression~\eqref{eq:summation} for the bare connected contribution to the singlet form factor
at vanishing momentum
for E300 at $t_{s,\rm min}=0.80\text{ fm}$ and D200 at $t_{s,\rm min}=0.90\text{ fm}$. The filled points indicate
the values that enter the fit; the quality of fit is indicated in the box on the figure.
}
\label{fig:summation_example}
\end{figure}

We report the numerical results of the form factor extraction on each ensemble as described in~\secref{sec:analysis}. 
In Tabs.~\ref{tab:coeff_s} and \ref{tab:coeff_singlet} we collect the results of
the simultaneous fits in $Q^2$ and $t_s$ for the $z$-expansion for the strange and singlet form factors, respectively.
We also report an example of the summation method applied to the connected contribution in~\figref{fig:summation_example}
for a single value of $Q^2$, which corresponds to the first step of the ``two-step'' procedure discussed in~\secref{subsec:connected}.

\begin{table}[h!]
\begin{tabular}{ccccccc}
\hline\hline
ID & $a_0$ & $a_1$ & $a_2$ & $\rho_{a_0,a_1}$ & $\rho_{a_0,a_2}$ & $\rho_{a_1,a_2}$ \\
\hline\hline
 H102 & $-0.0500(43)$ & $0.057(22)$ & $0.091(59)$ & $-0.57966$ & $0.20849$ & $-0.84444$  \\
 H105 & $-0.0417(69)$ & $0.070(34)$ & $-0.040(61)$ & $-0.79042$ & $0.35483$ & $-0.65399$  \\
 C101 & $-0.0415(54)$ & $0.058(27)$ & $0.046(56)$ & $-0.66147$ & $0.00103$ & $-0.56508$  \\
 N101 & $-0.0534(45)$ & $0.095(29)$ & $0.045(81)$ & $-0.56226$ & $0.31775$ & $-0.91011$  \\
 S400 & $-0.0524(68)$ & $0.064(28)$ & $-0.0009(500)$ & $-0.70915$ & $0.03087$ & $-0.47276$  \\
 N451 & $-0.0462(54)$ & $0.053(23)$ & $0.054(55)$ & $-0.59169$ & $-0.00101$ & $-0.66223$  \\
 D450 & $-0.0501(73)$ & $0.095(45)$ & $0.013(95)$ & $-0.77778$ & $0.33501$ & $-0.69722$  \\
 N203 & $-0.0547(47)$ & $0.114(35)$ & $0.023(92)$ & $-0.66042$ & $0.33043$ & $-0.87438$  \\
 N200 & $-0.0451(42)$ & $0.066(26)$ & $0.049(68)$ & $-0.62386$ & $0.27643$ & $-0.84611$  \\
 D200 & $-0.0405(58)$ & $0.076(29)$ & $-0.031(50)$ & $-0.83238$ & $0.20169$ & $-0.50324$  \\
 E250 & $-0.0332(88)$ & $0.037(47)$ & $-0.020(41)$ & $-0.87477$ & $0.16563$ & $-0.33639$  \\
 N302 & $-0.0413(88)$ & $0.052(27)$ & $0.038(53)$ & $-0.65772$ & $-0.08478$ & $-0.52741$  \\
 J303 & $-0.0378(81)$ & $0.063(31)$ & $0.019(64)$ & $-0.62573$ & $0.18523$ & $-0.65222$  \\
 E300 & $-0.0444(87)$ & $0.084(38)$ & $0.038(81)$ & $-0.63405$ & $-0.13794$ & $-0.54834$  \\
\hline\hline
\end{tabular}
\caption{Results for the coefficients of the $z$-expansion for the strange form factor $G_A^{s}(Q^2)$ for each ensemble, as well as their correlations $\rho$.}
\label{tab:coeff_s}
\end{table}

\begin{table}[h!]
\begin{tabular}{ccccccc}
\hline\hline
ID & $a_0$ & $a_1$ & $a_2$ & $\rho_{a_0,a_1}$ & $\rho_{a_0,a_2}$ & $\rho_{a_1,a_2}$ \\
\hline\hline
 H102 & $0.420(21)$ & $-0.43(16)$ & $-0.30(49)$ & $-0.55475$ & $0.24112$ & $-0.87515$  \\
 H105 & $0.482(51)$ & $-1.27(47)$ & $2.4(1.4)$ & $-0.67007$ & $0.40437$ & $-0.90007$  \\
 C101 & $0.364(28)$ & $-0.02(26)$ & $-1.94(81)$ & $-0.58428$ & $0.28658$ & $-0.86096$  \\
 N101 & $0.409(22)$ & $-0.41(15)$ & $0.35(48)$ & $-0.58757$ & $0.31939$ & $-0.86613$  \\
 S400 & $0.456(31)$ & $-0.61(25)$ & $0.007(819)$ & $-0.55464$ & $0.21355$ & $-0.87501$  \\
 N451 & $0.403(23)$ & $-0.16(17)$ & $-1.44(51)$ & $-0.57789$ & $0.22473$ & $-0.83573$  \\
 D450 & $0.366(36)$ & $-0.39(29)$ & $0.29(91)$ & $-0.68929$ & $0.36487$ & $-0.83337$  \\
 N203 & $0.332(25)$ & $0.04(22)$ & $-1.57(66)$ & $-0.59985$ & $0.28622$ & $-0.87190$  \\
 N200 & $0.397(34)$ & $-0.32(30)$ & $-0.18(95)$ & $-0.58087$ & $0.29409$ & $-0.88870$  \\
 D200 & $0.433(36)$ & $-0.93(33)$ & $0.55(98)$ & $-0.64949$ & $0.35969$ & $-0.87590$  \\
 E250 & $0.388(60)$ & $-0.51(62)$ & $0.7(1.9)$ & $-0.73652$ & $0.46906$ & $-0.89671$  \\
 N302 & $0.443(37)$ & $-0.51(26)$ & $-0.42(82)$ & $-0.50539$ & $0.15406$ & $-0.87375$  \\
 J303 & $0.362(40)$ & $-0.35(31)$ & $0.53(93)$ & $-0.62964$ & $0.34276$ & $-0.86904$  \\
 E300 & $0.354(45)$ & $-0.26(35)$ & $-1.52(97)$ & $-0.67212$ & $0.35520$ & $-0.86714$  \\
\hline\hline
\end{tabular}
\caption{Results for the coefficients of the $z$-expansion for the singlet form factor $G_A^{u+d+s}(Q^2)$ for each ensemble, as well as their correlations $\rho$.}
\label{tab:coeff_singlet}
\end{table}

\end{document}